\documentclass[a4paper, a4paperamsfonts, amssymb, amsmath,showkeys, nofootinbib, twoside]{revtex4-2}
\usepackage[english]{babel}
\usepackage[utf8]{inputenc}
\usepackage[colorinlistoftodos, color=green!40, prependcaption]{todonotes}
\usepackage[pdftex, pdftitle={Article}, pdfauthor={Author}]{hyperref} 
\usepackage[T1]{fontenc}
\usepackage{lmodern}
\usepackage{amsmath}
\usepackage{amssymb}
\usepackage{xcolor}
\usepackage{graphicx}
\usepackage{caption}
\usepackage{subcaption}
\usepackage{float}
\usepackage{esint}
\usepackage{dcolumn}
\usepackage{babel}
\usepackage{csquotes}
\usepackage{color}
\usepackage{slashed}
\usepackage{simplewick}
\usepackage{hyperref}
\hypersetup{
    colorlinks=true,
    citecolor=blue,
    filecolor=green,
    linkcolor=purple,
    urlcolor=red,
}
\usepackage{slashed}
\usepackage{hyperref}
\hypersetup{colorlinks,breaklinks,
			citecolor=[rgb]{0,0.0,1.0},
            urlcolor=[rgb]{0.0,0.0,1.0},
            linkcolor=[rgb]{0,0.5,0.9}}

\newcommand{\beq}{\begin{equation}}
\newcommand{\eeq}{\end{equation}}
\newcommand{\bea}{\begin{eqnarray}}
\newcommand{\eea}{\end{eqnarray}}

\begin{document}

\title{\textbf{Regularized Black Hole Solution from a New String Cloud Source}}

\author{Celio R. Muniz}
\email{celio.muniz@uece.br}
\affiliation{Universidade Estadual do Cear\'a, Faculdade de Educa\c c\~ao, Ci\^encias e Letras de Iguatu, 63500-000, Iguatu, CE, Brazil.}
\author{Jonathan Alves Rebou\c cas}\email{jalvesreboucas@ifce.edu.br} 
\affiliation{Universidade Estadual do Cear\'a, Faculdade de Educa\c c\~ao, Ci\^encias e Letras de Iguatu, 63500-000, Iguatu, CE, Brazil.}
\affiliation{Instituto Federal de Educação Ciências e Tecnologia do Ceará (IFCE), Iguatu, Brazil}
\author{Leonardo Tavares de Oliveira}\email{leonardo.tavares@uece.br} \affiliation{Universidade Estadual do Cear\'a, Faculdade de Educa\c c\~ao, Ci\^encias e Letras de Iguatu, 63500-000, Iguatu, CE, Brazil.}
\author{Francisco Tiago Barboza Sampaio}\email{tiago.barboza@uece.br} \affiliation{Universidade Estadual do Cear\'a, Faculdade de Educa\c c\~ao, Ci\^encias e Letras de Iguatu, 63500-000, Iguatu, CE, Brazil.}
\author{Francisco Bento Lustosa}\email{chico.lustosa@uece.br} \affiliation{Universidade Estadual do Cear\'a, Faculdade de Educa\c c\~ao, Ci\^encias e Letras de Iguatu, 63500-000, Iguatu, CE, Brazil.}

\begin{abstract}
We construct a new family of regular black hole solutions supported by the novel Letelier-Alencar string cloud and regularize it through a rational Dagum-type distribution. The regulator smooths the matter profile and ensures finite curvature invariants, yielding a geometry that interpolates between a string-cloud exterior and an anti-de Sitter core. We analyze the energy conditions, identifying where the null, weak, dominant, and strong conditions hold or fail across the core and exterior, given the regularity of the solution. The parameter space for horizon formation is mapped and the thermodynamic properties -- mass, Hawking temperature, entropy, and heat capacity -- are derived; notably, the entropy depends only on the regularization scale while the string parameter modifies temperature and heat capacity. Employing R\'enyi's non-extensive entropy and the approach of topological thermodynamics, we show that the non-extensive deformation stabilizes the system and eliminates the standard phase transition, leaving only a single thermodynamically stable black hole state. Finally, we compute the shadow radius and derive constraints compatible with the current Event Horizon Telescope bounds for Sgr~A* and M87*.
\end{abstract}
\maketitle

\section{Introduction}

The Theory of General Relativity (GR) has been successfully tested in a wide range of scales, from corrections to solar system dynamics, passing through a large number of astrophysical tests \cite{Berti_2015, LIGOScientific:2016aoc, LIGOScientific:2020ibl, akiyama2019first, EventHorizonTelescope:2022wkp} and going up to the largest scales in cosmology being the foundational background of the Standard $\Lambda$CDM model \cite{Planck2018,Ishak2019}. One of its most studied predictions is the formation of singularities through gravitational collapse, regions in space-time where the gravitational field tends to infinity and all physical descriptions breakdown \cite{Penrose1965, Hawking1966}. In the early Universe, first-order phase transitions with vacuum bubble collisions can also lead to the formation of primordial black holes through the collapse of false-vacuum bags \cite{etde_20498414,Khlopov1999}. This has motivated a large effort in the direction of developing modified or quantum theories of gravity that do not allow the existence of singularities \cite{rovelli2004quantum,Ashtekar2022, Bambi2023}. Penrose's cosmic censorship conjecture would preclude any direct observations of those singularities due to the formation of horizons \cite{Penrose1969pc}. This conjecture posits that gravitational singularities from collapse are hidden within Black Holes (BHs) by event horizons, preventing the emergence of "naked" singularities. From that point on the study of BHs has become one of the most explored subjects of modern physics, both through theoretical work focused on understanding its fundamental nature \cite{Wald:1995yp} and through phenomenological and observational work aimed at testing GR and its possible extensions \cite{RomeroVila2014}. The subject of BH thermodynamics, in particular, provides a wide variety of possible approaches that can possibly link the microscopic and fundamental structure around and inside the horizon to observable macroscopic features that can lead to testable predictions \cite{Bekenstein:1973ur,wald2001thermodynamics}. One of the most astonishing developments in this area has been the derivation of the Bekenstein-Hawking entropy from microscopic quantum dual states in the context of AdS/CFT providing another strong link between BH thermodynamics and quantum gravity theories \cite{Almheiri2021, Maldacena2023}.

Although the existence of astrophysical BHs at this point is largely taken as a fact due to the incredible fit between models and observations of gravitational waves and BH shadows \cite{LIGOScientific:2016aoc, LIGOScientific:2020ibl, akiyama2019first, EventHorizonTelescope:2022wkp}, the formation of singularities is not thought to be as inevitable as it once was \cite{Ashtekar2022, Bambi2023, NOVELLO2008}. The search for regular BH solutions has become one of the most important topics in recent literature, as they can be used to test indirect consequences of quantum and modified gravity theories \cite{Ayn-Beato2005,Fernandes:2020nbq,Ishibashi2021}. Although the singularity theorems developed by Penrose and Hawking indicate that divergences are inevitable under physically reasonable energy conditions, it is widely believed that the fundamental theory of gravitation should remain regular in high curvature regimes, in which quantum effects or nonlinear corrections become relevant.

In this context, the construction of regular black holes (RBHs) offers a particularly useful alternative to eliminate singularities, introducing smooth energy and momentum distributions that produce finite curvature invariants throughout spacetime. Early models, such as those proposed by  \cite{Bardeen1973,dymnikova1992vacuum,hayward2006formation}, replaced the central singularity with a de Sitter nucleus (dS) characterized by a finite effective cosmological constant, recovering the Schwarzschild vacuum in the asymptotic limit. Later developments connected these regular geometries to nonlinear electrodynamics (NLED) \cite{Ayn-Beato1998, Frolov:2016pav}, noncommutative geometry \cite{nicolini2009noncommutative} and modified gravity theories \cite{Nojiri2017, NunesdosSantos:2025alw}, establishing a broad phenomenological framework in which curvature regularization is associated with an effective matter distribution or quantum corrections of gravitation. These are only some of the approaches found in the literature to obtain regular BH solutions, but as long as they are fundamentally well motivated, the only way to access their validity is to derive their phenomenological consequences and test their consistency with observations. Broadly speaking, one can try to divide the approaches in three camps: i) solutions obtained from fundamental modifications of gravitational theories, possibly due to quantum effects; ii) solutions obtained from forcing regularity and reconstructing Einstein's field equations, and iii) solutions obtained from exotic of unusual matter source distributions or non-minimal couplings motivated by fundamental physics.

Among the physically motivated models of matter sources the string cloud case \cite{letelier1979clouds, letelier1981fluids} stands out due to its connection with the possibility that the fundamental entities of nature are string-like which has lead to the historic and still ongoing development of String Theory \cite{Kiritsis2019}.  A further motivation comes from the possibility of explaining galactic curves usually associated with Dark Matter (DM) by considering the presence of string fluids with an appropriate equation of state parameter \cite{Soleng:1993yr}. In \cite{Capozziello:2006ij} it was shown that a fluid of strings like the one proposed by Letelier could form part of the DM budget of our universe, as well as offering a Unified Dark Energy (UDE) scenario that could solve the cosmological constant and the coincidence problems \cite{MARTIN2012, Velten2014}. In addition, regular black hole remnants with de Sitter interiors have been proposed as heavy dark matter candidates, forming graviatoms that probe possible inhomogeneities of the early Universe \cite{Dymnikova2015}. Considering the current challenges facing the standard cosmological model, like the Hubble tension and possible variations of Dark Energy's equation of state parameter being indicated by DESI, studying further consequences of other matter source models that are fundamentally motivated is of particular importance. In Letelier's approach, the gravitational field is generated by a spherically symmetric distribution of one-dimensional strings radiating from the origin, which modifies the Schwarzschild potential through a constant term proportional to the string parameter $a$. However, Letelier's solution still presents a curvature singularity at the origin, which prevents a completely self-sustaining physical interpretation. Recently, several works have sought to regularize the singular structure of Letelier-type geometries using Dymnikova-type exponential regulators \cite{dymnikova1992vacuum, Dymnikova2004, NunesdosSantos:2025alw,Kar:2025phe}.  These regulators smoothly distribute the energy density around a finite scale $r_0$, producing space-times with de Sitter kernels and finite curvature invariants. Another recent development has been the proposal of a new cloud of string solution that includes a ``magnetic-like'' contribution to the energy momentum tensor that leads to a modified BH solution with a nonlinear dependence on string length $\ell_s$ and string coupling $g_s$ coupled to a hypergeometric function of $r^4/\ell_s^4$ \cite{ALENCAR2025102031}. This model provides a more complete physical description beyond Letelier's initial approach as its dependence on the string coupling can be viewed as encoding string backreaction in the gravitational field at tree level. This solution, which we will refer to as Letelier-Alencar solution, remains singular at the origin and has some limitations when dealing with evaporation up to the scale where quantum backreaction effects should become relevant but serves as an interesting effective extension to the original cloud of strings model and recovers it in the point-particle limit \cite{ALENCAR2025102031}. A natural next step would be to apply the Dymnikova-type regulators that can be viewed as connected to quantum vaccuum corrections \cite{dymnikova1992vacuum, Alencar:2023wyf, Estrada:2023pny, NunesdosSantos:2025alw} and study the regularity of the solutions. However, as we shall demonstrate, when combined with the hypergeometric correction that generalizes the original potential of the string cloud , the exponential damping becomes insufficient to eliminate all divergences, resulting in residual linear terms in the curvature near the origin. This indicates that standard exponential regularization, while successful for the Schwarzschild and Letelier solutions, does not provide complete regularization when higher-order corrections are incorporated into the string cloud model. 

To address these limitations, we introduce a new family of regular black hole solutions sourced by the novel Letelier-Alencar string cloud whose matter distribution is governed by a Dagum-type \cite{dagum2006wealth} rational regulator. This function smooths the energy density near the core, guarantees the finiteness of curvature invariants, and preserves the expected asymptotic behavior. The use of a rational regulator is physically motivated: similar profiles arise in nonlinear electrodynamic models of regular black holes, where they replace the central singularity with a finite-density core \cite{Balart2014}. Here, the Dagum regulator plays the same role, providing a smooth transition between the regular interior and the exterior string-cloud regime. The resulting spacetime interpolates between a Letelier-dominated exterior and an anti--de Sitter (AdS) vacuum at the origin, whose curvature radius is determined by the regularization scale $r_{0}$ and the string $a$ parameters. The analysis of the parameter space identifies the domains in which horizons form, as well as the critical boundaries where the horizon degenerates, linking macroscopic and microscopic regimes through the interplay between $a$, $r_{0}$, and the mass $M$.

The most common way to study the link between microscopic and macroscopic aspects of BH solutions is through their thermodynamic behavior. As already mentioned, since the first works of Bardeen and Hawking in the 1970's the study of BH thermodynamics has become a traditional way of understanding the stability and possible evolution of those objects through evaporation \cite{Bardeen1973, wald2001thermodynamics}. Beyond that, in this context one can study its phase space evolution, discuss phase transitions between different states and consider possible corrections to the standard laws of BH thermodynamics that include thermodynamic pressure from the cosmological constant \cite{Kastor_2009, Rehan2024}. Considering the importance of topology and critical points in the study of classical thermodynamic systems, an approach that has gained a lot of interest in the recent literature consists of using topological charges to understand BH stability, phase transitions and classify solutions in a general manner \cite{Wei:2021vdx, wei2022black}. BH solutions with extended phase spaces present a rich structure of phase transitions, the most widely found transition occurs between small and large solutions in a way that has been compared with the liquid-gas transition of Van de Waals fluids \cite{MANN2012, Wei:2015iwa, Wei:2019uqg, WEI2021}. As the system's temperature is lowered, a coexistence curve forms in the pressure-temperature plane, delineating the states where small and large black hole phases can exist simultaneously. This curve originates at a critical point and terminates at the origin of the diagram. In the vicinity of this critical point, the system exhibits universal critical behavior, whose properties are accurately described by mean field theory. While the specific shapes of phase diagrams can vary between different systems, the emergence of a critical point is a universal and fundamental feature. We employ this approach to study our regularized Letelier-Alencar string cloud solution and discuss how the two phases emerge.

The analysis of BH thermodynamics can be further enriched by the use of generalized entropies that go beyond the extensive Boltzmann-Gibbs traditional definition \cite{Renyi:1959pbs,Tsallis:1987eu, Tsallis2013}. Beyond the fact that the usual Bekenstein-Hawking formula is explicitly inconsistent with the usual relation $S \propto L^3$, modern developments in BH physics indicate that quantum gravitational effects might be connected to entanglement giving further motivation to deal with nonadditive definitions of entropies like the ones proposed by Tsallis or its Renyí-like version \cite{Biro:2013cra,Czinner:2015eyk, Czinner:2017tjq}. It has been shown that for any viable nonadditive entropy, there must exist a transformation—a "formal logarithm" denoted as $L(S)$—that converts it into an additive quantity ($L(S_{1 2}) = L(S_1) + L(S_2))$. This allows for a consistent definition of temperature via $1/T = \partial L(S(E))/ \partial E$. The Bekenstein-Hawking entropy is a natural example of a nonadditive entropy, following the rule $S_{1 2} = S_1 + S_2 + 2\sqrt{(S_1 S_2)}$. For more general cases with a non-zero parameter $\lambda$, this framework leads directly to the Rényi entropy. Specifically, when the original entropy follows the Tsallis composition rule $S_{1 2} = S_1 + S_2 + \lambda S_1 S_2$, its formal logarithm is precisely the Rényi entropy: $S_R = L(S_T)$. Both the Tsallis and Rényi formulas generalize standard thermodynamics and reduce to the conventional Boltzmann-Gibbs entropy in the limit where the non-extensivity parameter $q$ approaches $1$ $(\lambda \rightarrow 0)$.

We investigate the thermodynamic properties of this regularized configuration in detail, including the dependence of mass, Hawking temperature, entropy, and heat capacity on the model parameters. It is observed that, although $a$ and $r_0$ modify the temperature and mass, the entropy depends exclusively on the regularization scale, which reveals a separation between the microscopic degrees of freedom associated with the regular core and those linked to the external string field. Using Rényi entropy formalism, we show that non-additive corrections stabilize the system and smooth the traditional Hawking–Page phase transition, in agreement with recent studies of black holes in non-extensive formalisms \cite{wei2022black, Okyay_2022}. Furthermore, applying the approach of topological defects developed by \cite{wei2022black}, we classify the thermodynamic phases by means of the topological charge $W$, demonstrating that non-extensive effects transform the standard two-phase structure (stable and unstable branches) into a single globally stable configuration. From a phenomenological point of view, the model allows us to look at observational data. By calculating the shadow radius and comparing it with recent measurements of Sgr A* and M87* obtained by the Event Horizon Telescope (EHT) collaboration\cite{akiyama2019first, vgn2025}, we establish constraints on the parameters $a$ and $r_0$. The results show that the presence of a regular core tends to reduce the shadow size compared to the Schwarzschild case, a characteristic consistent with other regular geometries \cite{Jiang_2024}. This fact reinforces the physical robustness of the model and suggests that black holes regularized by string clouds can serve as effective descriptions of compact astrophysical objects. In summary, this work presents a regularized black hole solution generated by a generalized string cloud.

The article is structured as follows: Section \ref{scr} reviews the classical string cloud model; Section \ref{rbhnsc} presents the new regularized metric and its geometric properties; Section \ref{ECond} discusses the energy conditions; Section \ref{tmd} analyzes the thermodynamic aspects; Section \ref{bhtd} presents the topological interpretation and non-extensive thermodynamics; Section \ref{sa} addresses the phenomenological analysis of the shadow; and finally, Section \ref{conclusion} brings together the conclusions and perspectives of the study.

\section{String Cloud revision}\label{scr}

The Letelier string cloud model \cite{letelier1979clouds} describes a spherically symmetric distribution of strings whose gravitational field modifies the Schwarzschild metric by a constant term proportional to the string parameter $a$ :
\begin{equation}
f(r) = 1 - \frac{2M}{r} - a.
\end{equation}
Although physically insightful, this model has a curvature singularity at $r = 0$. Recent works have shown that such singularities can be removed by introducing exponential regulators of Dymnikova type \cite{NunesdosSantos:2025alw}, which smoothly distribute the matter content around a finite radius $r_0$, producing regular black holes with de Sitter-like interiors.

In a recent work, Alencar et al. \cite{ALENCAR2025102031} proposed an extension of the Letelier model that preserves its fundamental structure while introducing new elements inspired by the string cloud theory.
Following their approach, starting from the Nambu-Goto action \cite{letelier1979clouds}, that is,
\begin{equation}
S_{NG} = \int \sqrt{-\gamma}\mathcal{M} d\lambda^0 d\lambda^1, 
\quad
\gamma_{AB} = g_{\mu\nu} \frac{\partial x^\mu}{\partial \lambda^A} \frac{\partial x^\nu}{\partial \lambda^B},
\end{equation}
where  $\mathcal{M}$ is a dimensionless constant that characterizes the string, \(\lambda_0\) and \(\lambda_1\) are timelike and spacelike parameters, respectively, and $\gamma$ is the determinant of the induced metric \(\gamma_{AB}\); the stress-energy tensor is written as
\begin{equation}
    T^\mu_\nu = \rho_p\frac{ \, \Sigma^{\mu\alpha}\Sigma_{\alpha\nu}}{8\pi \sqrt{-\gamma}},
\end{equation}
 using the definitions 
 \(\Sigma^{\mu\nu} = \epsilon^{AB} \frac{\partial x^\mu}{\partial \lambda^A} \frac{\partial x^\nu}{\partial \lambda^B}\),  \(\gamma = \frac{1}{2} \Sigma^{\mu\nu} \Sigma_{\mu\nu}
\), where $\rho_p$ is the proper density and \(\Sigma^{\mu\nu}\) satisfies the following equations
\begin{equation}
    \nabla_\mu (\rho_p \Sigma^{\mu\nu}) = 0, \quad \Sigma^{\mu\beta}\nabla_\mu \left[\frac{\Sigma_\beta{}^\nu}{\sqrt{-\gamma}}\right] = 0.
\end{equation}

For spherically symmetric solutions, the authors introduced an additional \enquote{magnetic-like} component,  \(\Sigma_{23}\), which also ensures 
$\gamma<0$, complementing the \enquote{electric-like} component, \(\Sigma_{01}\), present in Letelier formulation.
This modification resulted in an stress–energy tensor of the form \(diag(-\rho,\rho,p,p)\), characterized by two constants and governed by a  unique equation of state.
Consequently, this generalization provides a new black hole solution, given by
\begin{equation}
\displaystyle f(r) = 1 - \frac{2M}{r} +  \frac{g_s^2 \ell_s^2}{r^2}\,
{}_2F_1\left(-\tfrac{1}{2}, -\tfrac{1}{4}; \tfrac{3}{4}; -\frac{r^4}{\ell_s^4}\right),\label{fr-geova}
\end{equation}
where \({}_2F_1\) is the Gaussian hypergeometric function, \(\ell_s\) is the string length and \(g_s\) the string coupling constant.
It was also observed that the solution exhibits two event horizons, defined by \(f(r)=0\), for \(g_s^2<1\).

On the other hand, for \(r \gg \ell_s\), the asymptotic behavior of the solution (\ref{fr-geova}) is
\begin{equation}
f(r) = 1 - \frac{2M}{r} - g_s^2 + \frac{g_s^2 \ell_s^4}{6r^4},
\end{equation}
which ensures that the solution is asymptotically flat. Moreover, in the limit \(\ell_s\to 0\), one recovers the Letelier solution, with \(a=g_s^2\).

\section{Regular Black Hole Sourced by the New String Cloud}\label{rbhnsc}

The Letelier-Alencar string cloud solution provides a more complete framework to study possible consequences of string backreaction into the spacetime structure \cite{ALENCAR2025102031}, but it remains a singular solution that might indicate that other effects might have to be taken into account. Here we first try to work with a more general modified form of the metric function, to be introduced to regularize the geometry while preserving the string cloud contribution, which can initially be written as
\begin{equation}
f(r) = 1 - \left[ \frac{2M}{r} - 
\frac{|a| r_0^2}{r^2} \,
{}_2F_1\!\left(-\tfrac{1}{2}, -\tfrac{1}{4}; \tfrac{3}{4}; -\frac{r^4}{r_0^4}\right)
\right]\!\left( 1 - e^{-r^3/r_0^3} \right),
\label{metric1}
\end{equation}
where the exponential factor introduces a natural cutoff at scale $r_0$, which seeks to ensure regularity near the origin and a smooth transition to the asymptotic region. This scale is made here equal to the characteristic length associated with the string cloud introduced in \cite{ALENCAR2025102031}. The exponential factor thus defines a natural cutoff scale at $r_0$ and regulates the spatial profile of the energy density of the string cloud. The factor containing the exponential term corresponds to a Dymnikova-like mass distribution (which is connected to quantum vaccum effects \cite{dymnikova1992vacuum}), which modifies the Schwarzschild geometry to produce regular de Sitter and Schwarzschild vaccua in the core and asymptotic regions, respectively. In our framework, for $r \gg r_0$, the metric (\ref{metric}) with this Dymnikova-like factor reduces to 
\begin{equation}\label{asympt}
f(r) \simeq 1 - \frac{2\widetilde{M}}{r} - |a|,
\end{equation}
introducing a correction to the black hole mass while recovering the standard Letelier string cloud behavior in the limit $r_0 \to 0$. The corrected black hole mass is given by  
\begin{equation}
\widetilde{M} = M - \frac{|a| r_0 \left[\Gamma\!\left(-\tfrac{1}{4}\right)\Gamma\!\left(\tfrac{3}{4}\right)\right]}{4\sqrt{\pi}} \approx M - 0.847\,|a|r_0,
\end{equation}
which shows that the presence of the scale $r_0$ slightly diminishes the effective mass of the configuration.

On the other hand, in the limit $r \to 0$, the metric coefficient takes the approximate form
\begin{equation}
f(r) \approx 1 + \frac{|a|\, r}{r_0} - \frac{2 M}{r_0^3}\, r^2.
\end{equation}
The geometry described by Eq.~(\ref{metric1}), while regular in its metric components, unfortunately leads to divergent curvature invariants in the origin due to the linear term in $r$. This indicates that the Dymnikova-type exponential regulator, while successful for the Schwarzschild and Letelier solutions, is insufficient to fully regularize the spacetime in the presence of the new string cloud. One could in principle consider extending the Dymnikova regulator to a quartic argument in the exponential, but there is no physical justification for such a modification, since higher-order powers are typically associated with higher-dimensional settings \cite{Estrada:2023pny}. We therefore introduce another regularization scheme based on a rational distribution function, which modifies the metric potential to a new form and successfully ensures the finiteness of all curvature invariants. 

Thus, instead of the Dymnikova-like factor, we will take the factor based on the Dagum distribution for both the mass and the string cloud, in such a manner that the metric potential becomes
\begin{equation}
f(r) = 1-\frac{2M(r)}{r}=1 - \left[ \frac{2M}{r} - 
\frac{|a| r_0^2}{r^2} \,
{}_2F_1\!\left(-\tfrac{1}{2}, -\tfrac{1}{4}; \tfrac{3}{4}; -\frac{r^4}{r_0^4}\right)
\right][1+(r_0/r)^c]^{-(c+1)\beta/c},
\label{metric}
\end{equation}
where we will choose $c=1$ and $\beta=2$. This reproduces the asymptotic behavior given in (\ref{asympt}), with the mass correction now expressed as
\begin{equation}
    \widetilde{M} = M -\frac{r_0 \Gamma \left(\frac{1}{4}\right) | a| }{2\Gamma \left(\frac{5}{4}\right)}+\frac{r_0\Gamma \left(-\frac{1}{4}\right) \Gamma \left(\frac{3}{4}\right) | a| }{4 \sqrt{\pi }} \approx M - 2.847\,|a|r_0.
\end{equation}
In contrast, near the core the metric takes the form
\begin{equation}
    f(r) \approx 1 + \frac{|a| r^2}{r_0^2},
\end{equation}
which corresponds to an anti--de Sitter vacuum, ensuring the regularity of all curvature invariants at the origin. We note that the chosen Dagum-type regulator represents a smooth, rational function that governs the spatial distribution of the string cloud energy density. The rational decay of the regularization factor is motivated by its proven efficacy in generating regular black holes from nonlinear electrodynamics (NLED) sources \cite{Balart:2014cga}. This regulator effectively ``smears'' the mass and string tension over a finite region of radius $r_0$, mimicking the introduction of a fundamental minimal length scale where classical gravity is modified \cite{Frolov:2016pav}. The specific choice of the exponents in the rational factor represents the minimal configuration required to keep all curvature invariants finite, yielding a mathematically consistent and physically motivated regular geometry, also inducing a transition to an anti-de Sitter core, a feature often linked to confining vacuum states in gauge/gravity duality. 

\begin{figure}[hb!]
    \centering 
    \includegraphics[width=0.496\textwidth]{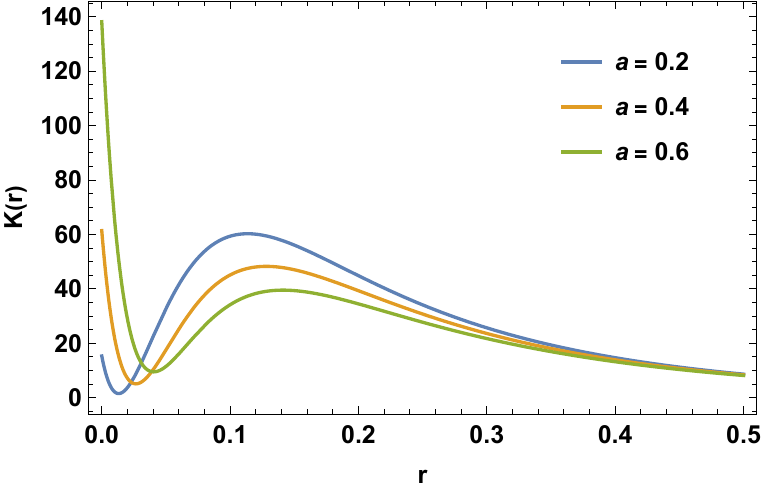}
     \includegraphics[width=0.496\textwidth]{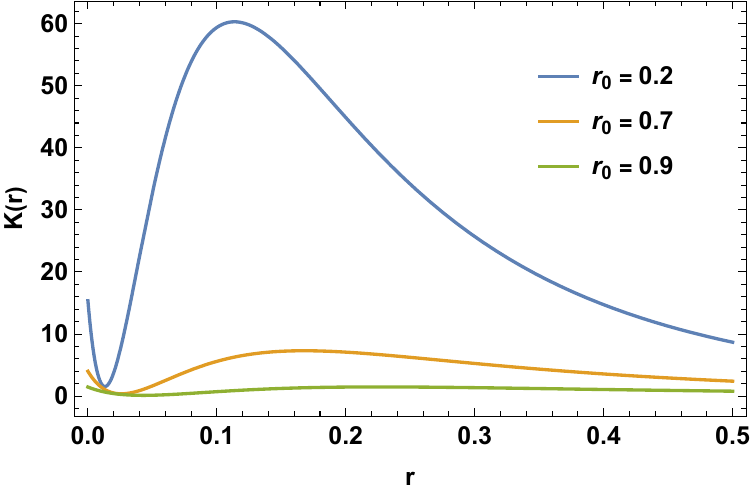}	
    \caption{Left panel: Kretschmann scalar as a function of $r$, varying the string parameter $a$, with $r_0=0.5$. Right panel: The same quantity, now varying the scale $r_0$, with $a=0.2$. It was considered $M=1.0$.}
    \label{kretsh}
\end{figure}

Let us now analyze the behavior of the curvature, specifically the Kretschmann scalar denoted by $K(r)$. It is defined as $K(r) = R_{\mu\nu\sigma\lambda}R^{\mu\nu\sigma\lambda}=(f'')^2 + \frac{4}{r^2}(f')^2 + \frac{4}{r^4}(1-f)^2$. In the approximation $r\to 0$, we have, for the metric potential of Eq. (\ref{metric})
\begin{equation}
    K(r)\approx \frac{24|a|^2}{r_0^4},
\end{equation}
confirming thus the regularity of the solution, as can be observed in Figure \ref{kretsh}, where the Kretschmann scalar, as a function of the radial coordinate, is depicted.

Interestingly, the metric potential (\ref{metric}) can be written in a sufficiently large region around the origin in the form of a Frolov-type metric:
\begin{equation}
    f(r)\approx 1-\frac{r^2(2mr-Q^2)}{l^2(2mr+Q^2)+r^4},
\end{equation}
where the parameters are defined as
\begin{align}
l&=\frac{r_0}{\sqrt{|a|}},\\[1pt]
q &= \left[ \frac{4M + 5|a|\,r_0}{2\,r_0\,(M + 2|a|\,r_0)} \right]^{1/6}, \\[1pt]
m &= \frac{(2M + 4|a|\,r_0)^{2/3}\,(4M + 5|a|\,r_0)^{1/3}}{4\,|a|\,r_0^{7/3}}.
\end{align}

This parametrization makes the analogy explicit with the class of regular black holes discussed by Frolov \cite{Frolov:2016pav}, whose core approaches an anti-de Sitter geometry. In fact, an expansion of the metric function around \(r=0\) yields
\begin{equation}
f(r)=1+\frac{r^2}{l^2}+O(r^3),
\end{equation}
indicating that the inner region is locally anti-de Sitter with an effective curvature radius \(l\) and an effective cosmological constant \(\Lambda_{\mathrm{eff}} = -3/l^2\). Consequently, both geometries converge toward the same AdS core in the limit \(r\to 0\). Deviations between the two metrics appear only at orders equal or greater than $\mathcal{O}(r^5)$, reflecting the distinct sources of matter involved, the present model being governed by the string cloud and regularization parameters \(a,r_0\), respectively. These higher-order corrections encode the fine structure of the regular core and influence phenomenologically relevant quantities such as the photon-sphere radius and the shadow size.

The source for the solution given by Eq. (\ref{metric}) is given by the density profile
\begin{equation}
   \rho(r)= \frac{| a|  \left[r_0^2(r+r_0) \sqrt{1+\frac{r^4}{r_0^4}} -4 r_0^3 \, _2F_1\left(-\frac{1}{2},-\frac{1}{4};\frac{3}{4};-\frac{r^4}{r_0^4}\right)\right]+8 M  r_0r}{8 \pi (r+r_0)^5},
\end{equation}
\begin{figure}[htp!]
    \centering 
    \includegraphics[width=0.48\textwidth]{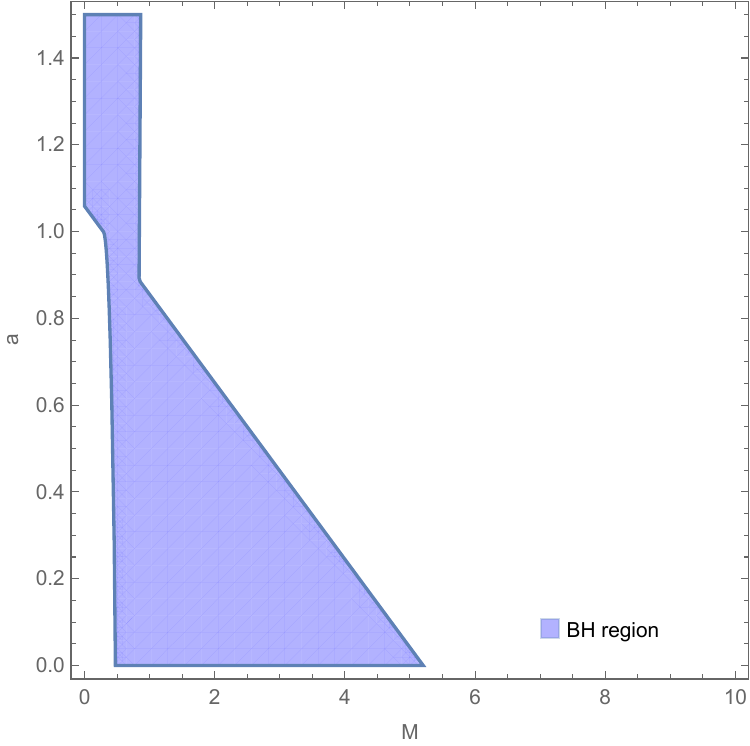}
    \includegraphics[width=0.48\textwidth]{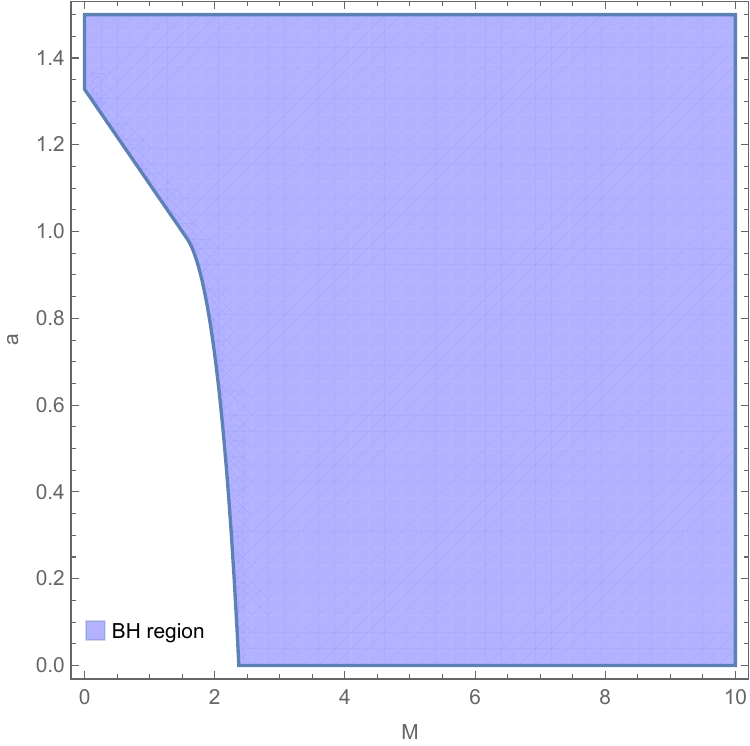}
    \caption{Parameter space $(M,a)$ of the regularized new string cloud black hole. The shaded region indicates the values of the parameters yielding black hole solutions with event horizons, bounded by the critical line where the horizon degenerates. Beyond this boundary, the spacetime describes no black hole. The regularization core radius is set to $r_0=0.1$ (left panel) and $r_0=0.5$ (right panel).}
    \label{fig:param_space}
\end{figure}
which is obtained from the mass function $M(r)$. In the vicinity of the origin, we have
\begin{equation}
\rho(r)\approx-\frac{3 | a| }{8 \pi r_0 ^2},
\end{equation}
which corresponds to a negative cosmological constant, and asymptotically 
\begin{equation}
\rho(r)\approx \frac{|a| }{8 \pi  r^2},
\end{equation}
which yields the density associated with the Letelier string cloud, as expected.

Regular black hole solutions arise from this source for specific parameter combinations, as illustrated in Fig.~\ref{fig:param_space}. The blue region represents the allowed values of the mass \(M\) and string cloud parameter \(a\) that yield real, positive roots of \(f(r) = 0\), corresponding to event horizons. The boundary of this region marks the critical transition between black holes and naked cores, as the regularization mechanism removes the central singularity through the smooth string cloud distribution. Moreover, smaller values of the regularization scale \(r_0\) favor the formation of microscopic black holes, while larger values of \(r_0\) extend the parameter domain toward macroscopic configurations, indicating that the regularization scale governs the transition between quantum and classical gravitational regimes.

\section{Energy Conditions}\label{ECond}

In static, spherically symmetric geometries supported by an anisotropic fluid with $T^{\mu}{}_{\nu}=\mathrm{diag}(-\rho,p_r,p_t,p_t)$, the pointwise energy conditions read
\begin{equation}
\begin{aligned}\label{EC}
&\text{NEC:}\quad &&\rho+p_r\ge 0,\quad \rho+p_t\ge 0,\\
&\text{WEC:}\quad &&\rho\ge 0\ \text{and NEC},\\
&\text{DEC:}\quad &&\rho - |p_r|\ge 0,\quad \rho - |p_t| \ge 0,\\
&\text{SEC:}\quad &&\rho+p_r+2p_t\ge 0.
\end{aligned}
\end{equation}
A characteristic algebraic feature in many regular black hole models is $T^t{}_t=T^r{}_r$ (equivalently $p_r=-\rho$), which arises, for example, in nonlinear electrodynamics (NED) sources. In such settings, the regularity of the center together with the existence of horizons entails a necessary violation of the strong energy condition (SEC) somewhere in the interior. This can be expressed by integrating the Einstein equations in a gauge
\begin{equation}
ds^2=-f(r)\,dt^2+f(r)^{-1}\,dr^2+r^2\,d\Omega^2,
\end{equation}
which yields an integral constraint on the Tolman mass $(m_T)$ between consecutive zeros of $f$,
\begin{equation}
\begin{aligned}
m_T[a,b]
&=4\pi\!\!\int_a^b\! dr\,r^2\,(T^k{}_k-T^t{}_t)\\
&=4\pi\!\!\int_a^b\! dr\,r^2\,(\rho+p_r+2p_t).
\end{aligned}
\end{equation}
In each static region under the horizon one finds $m_T<0$, hence $\int r^2(\rho+p_r+2p_t)\,du<0$ and SEC violation there; complementary constraints apply in nonstatic regions. Therefore, for regular black holes with a regular center and at least one horizon, SEC violation in static interior domains is unavoidable, while the other conditions may or may not be violated depending on the degree of anisotropy \cite{Zaslavskii2010}.

Within Einstein–NED models this structure is realized explicitly. If the NED Lagrangian $L(F)$, where $F \equiv F_{\mu\nu}F^{\mu\nu}$ is the electromagnetic invariant, retains the Maxwell weak–field limit $L\!\sim\! F$ as $F\!\to\!0$, then globally regular, static, spherically symmetric solutions with nonzero \emph{electric} charge are ruled out: a regular center requires $L\!\to\!\text{const}$ as $F\!\to\!\infty$, which is naturally achieved by \emph{magnetic} configurations. The resulting magnetic regular black holes and solitons satisfy $T^t{}_t=T^r{}_r$, develop a de~Sitter–type core with $p_r=p_t=-\rho$, and can obey the weak energy condition (WEC) while remaining everywhere regular \cite{Bronnikov2001}. Electrically charged regular solutions become possible if the NED dynamics departs from the Maxwell limit near the core; in such cases one can preserve the WEC and obtain a vacuum–like central equation of state that provides a finite self–energy cutoff, with one or two horizons depending on parameters \cite{Dymnikova2004}. 

A constructive route consists in prescribing a regular mass function $M(r)$ with $dM/dr\ge 0$ and the algebraic constraint $T^t{}_t=T^r{}_r$, which guarantees $p_r=-\rho$ and allows $p_t$ to be chosen so that the WEC (and often the DEC) holds everywhere while the SEC fails only in a compact inner region, precisely as dictated by the Tolman–mass integral \cite{Balart2014}. Earlier exact NED solutions by Ay\'on-Beato  and Garc\'ia  \cite{Ayn-Beato2005} exemplify the same pattern: for suitable choices of the parameters that control the NED “multipole’’ moments, curvature invariants remain finite, horizons exist for bounded charge-to-mass ratio, asymptotics are Reissner–Nordström-like, and the WEC can be enforced, whereas the center behaves as an AdS–like vacuum rather than a Maxwellian core \cite{Ayn-Beato1998,Ayn-Beato2005}.

From an effective perspective, renormalization–group (RG)–improved black holes implement running couplings $G\!\to\!G(k)$, with a diffeomorphism–invariant scale identification $k=k(\text{curvature})$. The resulting quantum–improved geometries replace the classical central singularity by a regular core whose local equation of state approaches a vacuum phase (Minkowski, de~Sitter, or anti–de~Sitter) depending on the identification scheme and parameters. Although the stress–energy is then “effective’’ rather than fundamental, the same logical outcome persists: regularity of the core correlates with SEC violation in the deep interior, while asymptotic regions recover the standard energy conditions as the running freezes \cite{Ishibashi2021}.

In summary, a coherent picture emerges across microscopic (NED) and effective (RG–improved) constructions. Regularity is achieved by a vacuum–like inner phase that tames curvature at $r\!=\!0$; this phase necessarily violates the SEC in static regions beneath the horizon, as quantified by the negative Tolman mass there. At the same time, many regular black hole families retain the WEC (and often the DEC) everywhere outside a compact core, either through magnetic NED realizations with Maxwell behavior at infinity \cite{Bronnikov2001} or via non–Maxwellian electric cores \cite{Dymnikova2004,Ayn-Beato1998,Ayn-Beato2005}, and can be engineered systematically through admissible mass functions \cite{Balart2014}. Quantum–improved solutions, while conceptually different, echo the same structural link between singularity resolution, vacuum–like cores, and interior SEC breakdown \cite{Ishibashi2021}.

In what follows we analyze the energy conditions for the present regular black hole surrounded by a string cloud, treating the string parameters $(r_{0},a)$ as \emph{control parameters} and tracking how the admissible domains of NEC/WEC/DEC/SEC evolve as they vary. From Einstein Field Equations $G_{\mu\nu}=\kappa\,T_{\mu\nu}$, with $T^{\mu}{}_{\nu}=\mathrm{diag}(-\rho,p_{r},p_{t},p_{t})$, one obtain
\begin{align}
\rho= - p_r =\frac{1-f(r)-r f'(r)}{\kappa\,r^{2}}\label{EFE1}\\
p_t=\frac{r f''(r)+2 f'(r)}{2\kappa\,r}\label{EFE2},
\end{align}
with $\kappa = 8\pi$, for geometric units.

From \eqref{EC}, \eqref{EFE1} and \eqref{EFE2}, we have WEC and SEC, for any $r$, explicitly given by

\begin{align}
\rho+p_r=0\\
\rho+p_t=\frac{2-2f+r^{2}f''}{2\kappa r^{2}}\\
\rho+p_r+2p_t=\frac{r f''+2 f'}{\kappa r}.
\end{align}

For the regularized string–cloud black hole, two limiting regimes are especially transparent. At spatial infinity one finds
\begin{equation}
\lim_{r\to\infty}\,p_t(r)=0,
\qquad
\lim_{r\to\infty}\,[\rho(r)+p_t(r)]=0,
\end{equation}
so the matter sector asymptotes to the canonical string–cloud profile with
$\rho\sim |a|/(8\pi r^{2})>0$, $p_r=-\rho$, and negligible tangential stress; the null/weak energy conditions are then satisfied asymptotically, in agreement with the large–$r$ behavior of string–cloud geometries reported in the literature. At the center the model approaches an AdS–like vacuum, and your limits give
\begin{equation}
\lim_{r\to 0}\,[\rho(r)+p_t(r)]=0,
\qquad
\lim_{r\to 0}\,[\rho(r)+p_r(r)+2p_t(r)]=\frac{3|a|}{4\pi r_{0}^{2}}>0,
\end{equation}
which implies $p_r(0)=p_t(0)=-\rho(0)$ with $\rho(0)<0$ and SEC/NEC satisfied  at the very center. This contrasts with many NED–based regular black holes—typically de Sitter–core solutions with $\rho(0)>0$ and $p_r=p_t=-\rho$, for which $\rho+p_r+2p_t=-2\rho(0)<0$ and the SEC is violated right at $r=0$ \cite{Bronnikov2001,Dymnikova2004,Balart2014}. In our case the sign flip (AdS core) explains the positive central limit above, while Zaslavskii’s integral constraint \cite{Zaslavskii2010} still mandates SEC violation somewhere in the \emph{static} interior between horizons; hence $p_t(r)$ must become negative in a finite radial layer away from $r=0$, reconciling the positive central limit with the required negative Tolman mass under the horizon. Altogether, the limits $r\to 0$ and $r\to\infty$ delineate the transition from a vacuum–like AdS core (SEC locally satisfied, $\rho<0$) to a string–cloud exterior (WEC/DEC satisfied, $p_t\to 0$), with the SEC–violating region confined to the intermediate region.

\begin{figure}[h!]
    \centering
    \includegraphics[width=0.49\linewidth]{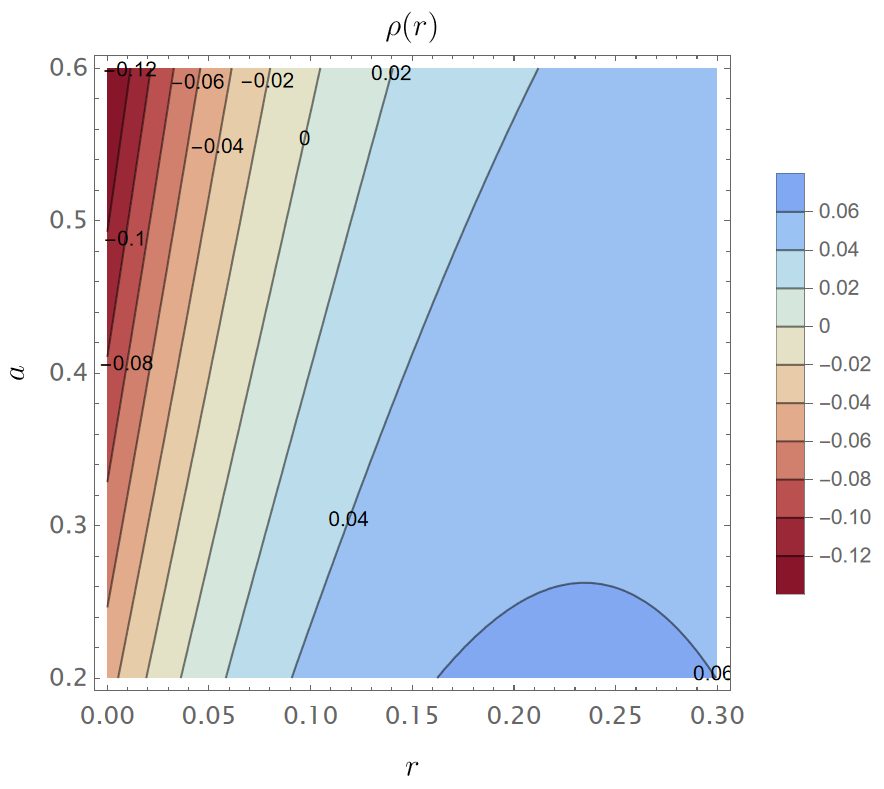}
    \includegraphics[width=0.49\linewidth]{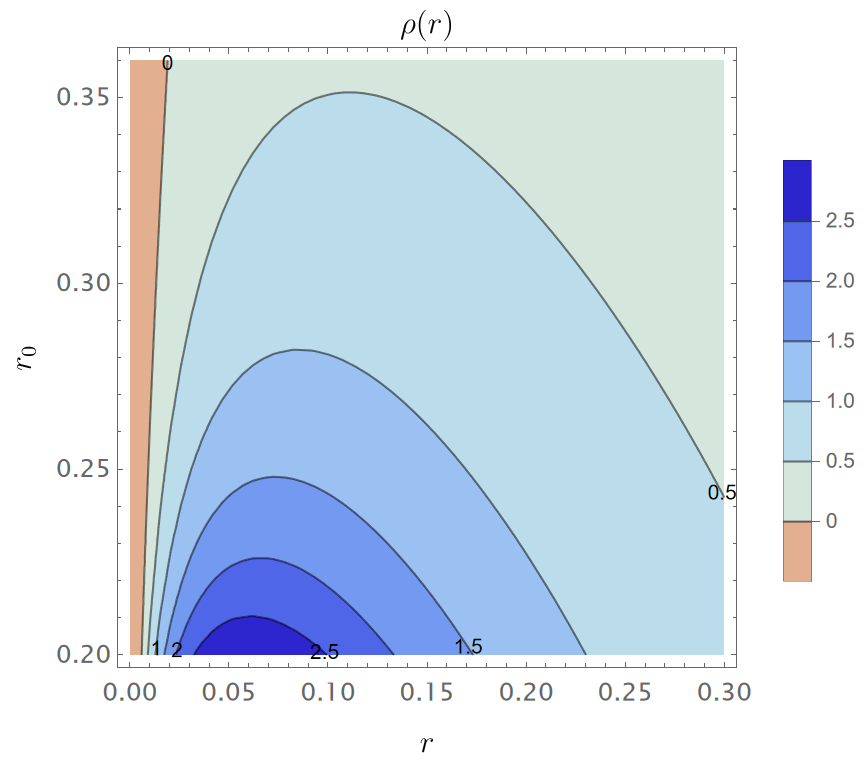}
    \caption{Contours of the energy density $\rho(r)$. In the {left} panel, we set $r_{0}=0.7$ and $M=1$; larger $a$ expands the inner region with $\rho<0$. In the {right} panel, we set $a=0.4$ and $M=1$; increasing $r_{0}$ moves the $\rho$ peak to larger $r$ and reduces its amplitude. The line $\rho=0$ marks the WEC threshold.}
    \label{fig:rhoEC}
\end{figure}

Figure~\ref{fig:rhoEC} shows how the string cloud controls the sign structure of the energy density. For fixed $r_{0}$ (left), increasing $a$ expands the $\rho<0$ domain and shifts the zero of $\rho$ to larger $r$, i.e., the AdS-like core deepens and the WEC-satisfying region retreats outward. For fixed $a$ (right), increasing $r_{0}$ pushes the positive-density peak to larger $r$ while lowering its amplitude, reflecting the smoothing role of the regularization scale. This pattern dovetails with the string-cloud asymptotics ($\rho\sim |a|/8\pi r^{2}>0$ at large $r$) and differs from canonical NED regular black holes, where de Sitter cores produce $\rho(0)>0$ \cite{Bronnikov2001,Dymnikova2004}; here the core is AdS-like ($\rho<0$) but the exterior still attains $\rho>0$, as in mass-function constructions that tune interior/exterior domains via model parameters \cite{Balart2014}. Combined with the integral constraint under horizons \cite{Zaslavskii2010}, the maps imply a SEC-violating band located between the negative-density core and the $\rho>0$ exterior, whose extent grows with $a$ and shrinks as $r_{0}$ increases.

\begin{figure}[h!]
    \centering
    \includegraphics[width=0.49\linewidth]{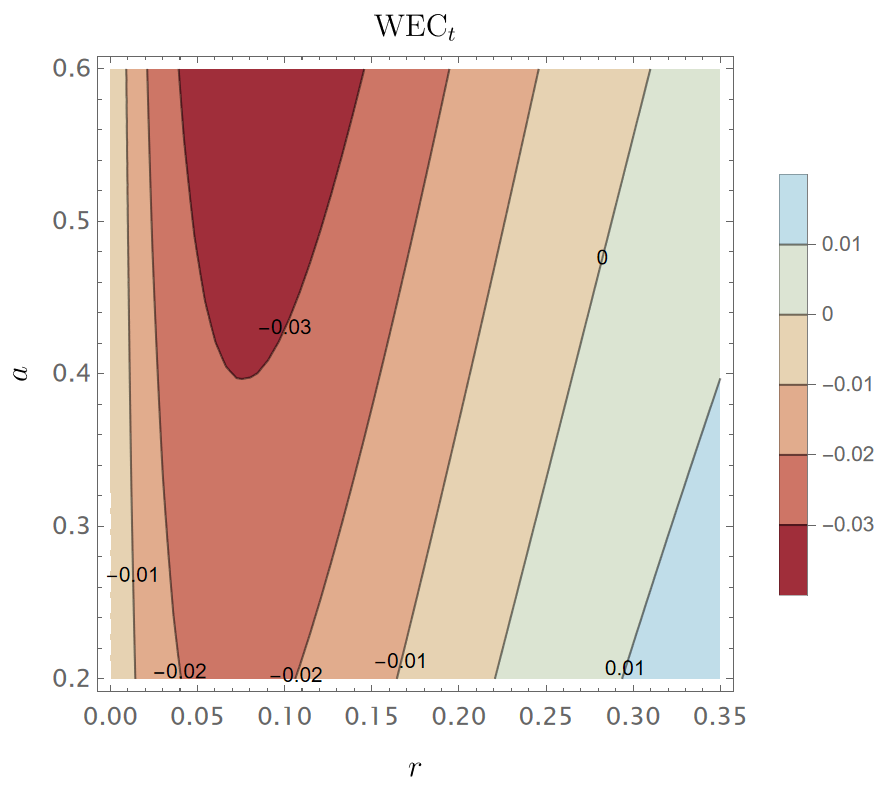}
    \includegraphics[width=0.49\linewidth]{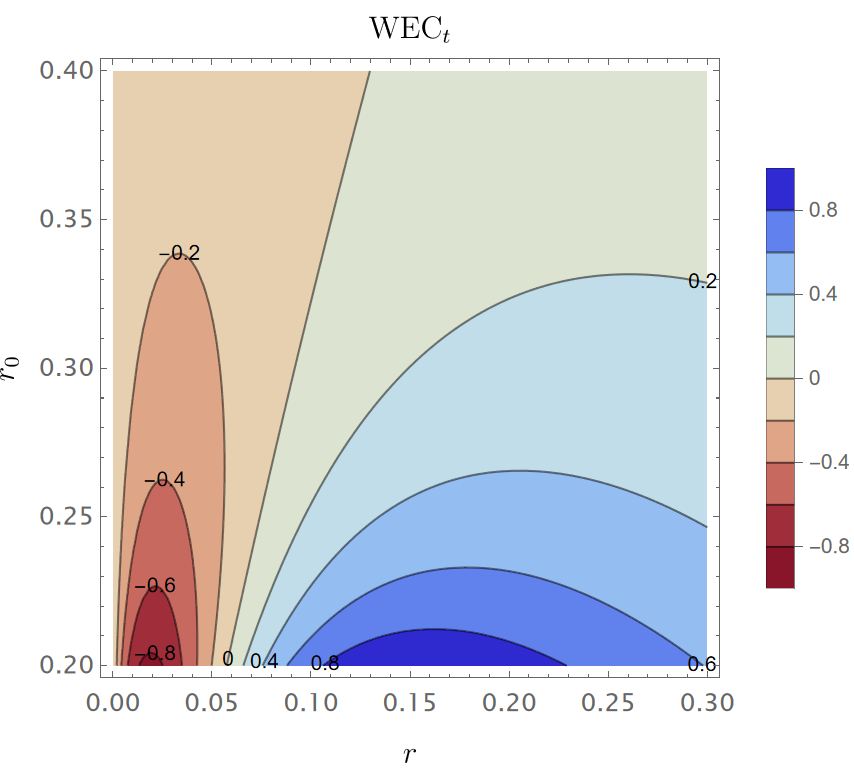}
    \caption{Contours of the tangential WEC ($\text{WEC}_t$). In the {left} panel, we set $r_{0}=0.7$ and $M=1$; larger $a$ drives $\rho+p_t$ more negative at fixed $r$ and shifts the $\rho+p_t=0$ curve outward. In the {right} panel, we set $a=0.4$ and $M=1$; increasing $r_{0}$ raises $\rho+p_t$ and moves its maximum to larger $r$. The line $\rho+p_t=0$ marks the boundary of the WEC$_t$ domain.}
    \label{fig:wec}
\end{figure}

Figure~\ref{fig:wec} maps the tangential WEC ($\mathrm{WEC}_t$), clarifying where the anisotropic sector preserves or violates the WEC/NEC. For fixed $r_{0}$ (left), increasing the string strength $a$ drives $\rho+p_t$ negative over a wider radial band and pushes the $\rho+p_t\!=\!0$ frontier outward; this enlarges the layer that accompanies the negative Tolman mass inside the horizons, in line with the expectation of an interior SEC-violating zone. For fixed $a$ (right), a larger regularization scale $r_{0}$ raises $\rho+p_t$ and shifts its maximum to larger $r$, shrinking the violation region and smoothing the transition to the string-cloud exterior where $p_t\!\to\!0$ and $\rho>0$. The central saturation $\rho+p_t\!\to\!0$ agrees with regular-core models \cite{Bronnikov2001,Dymnikova2004}, but unlike the typical NED case—where $\rho(0)>0$ and $\rho+p_t$ tends to stay nonnegative near the core—our AdS core favors a negative $\mathrm{WEC}_t$ band away from $r\!=\!0$, whose extent grows with $a$ and diminishes with $r_{0}$, consistent with the mass-function diagnostics and Zaslavskii’s integral constraint \cite{Balart2014,Zaslavskii2010}.

\begin{figure}[h!]
    \centering
    \includegraphics[width=0.49\linewidth]{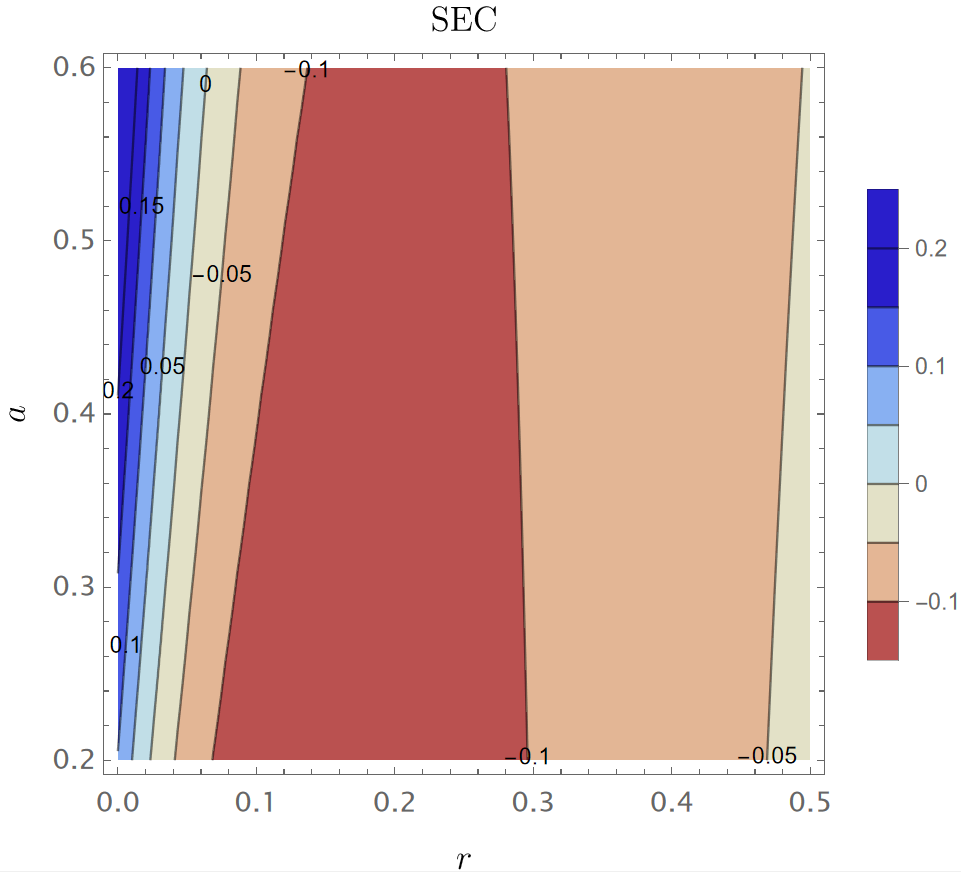}
    \includegraphics[width=0.49\linewidth]{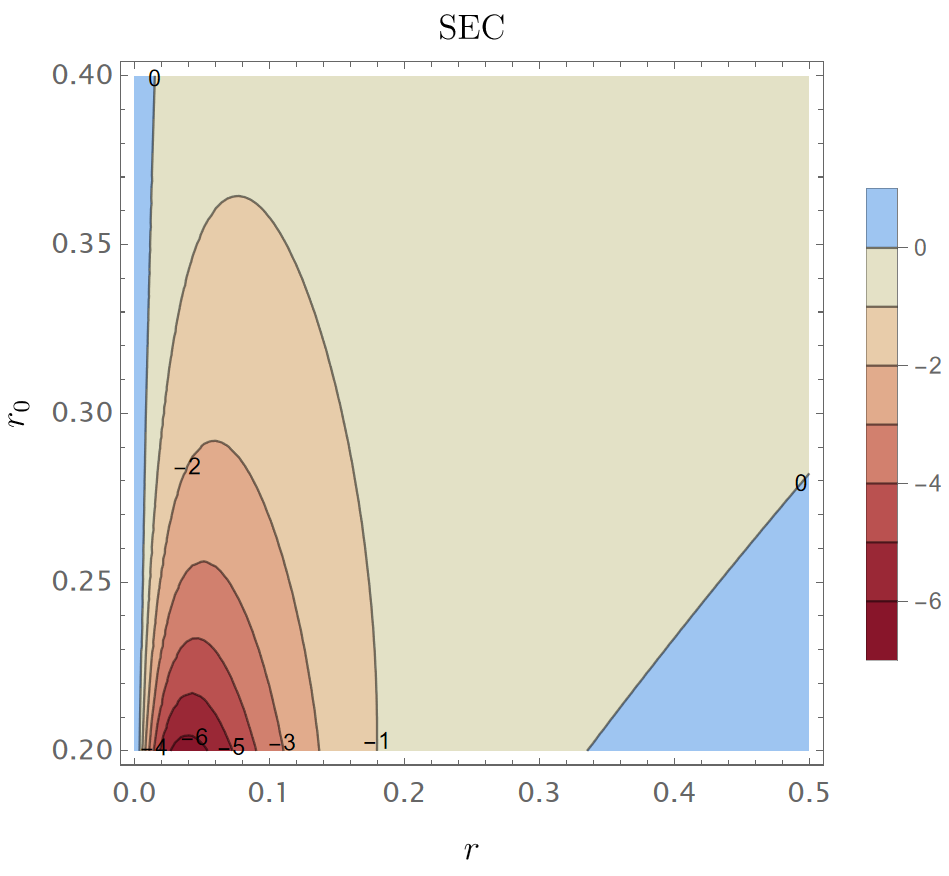}
    \includegraphics[width=0.5\linewidth]{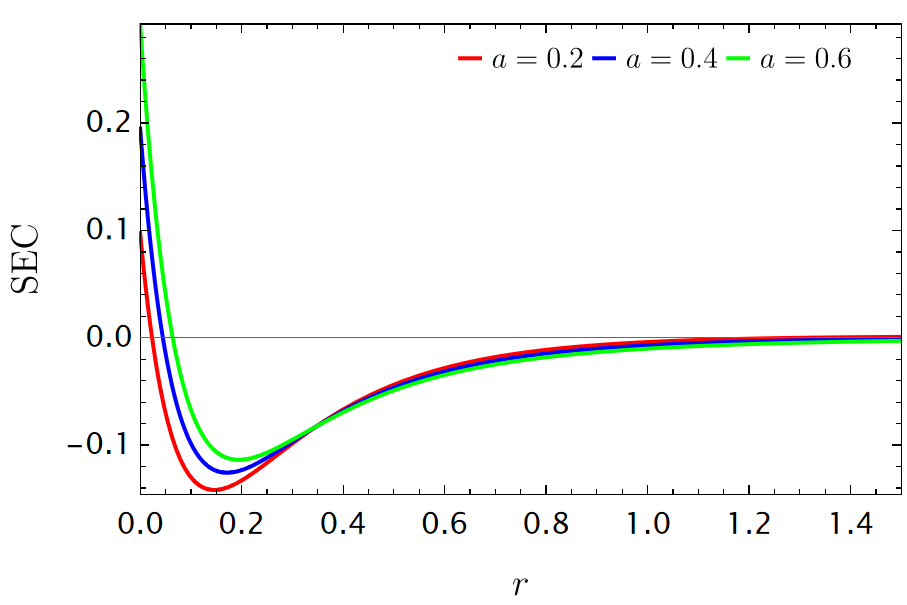}
    \caption{Contours of the SEC. In the {top-left} panel, we set $r_{0}=0.7$ and $M=1$; larger $a$ widens the negative–SEC region and shifts the $\mathrm{SEC}=0$ curve outward. In the {top-right} panel, we set $a=0.4$ and $M=1$; increasing $r_{0}$ raises $\mathrm{SEC}$ and moves its zero to larger $r$, shrinking the negative band. In the {bottom} panel, the radial profiles of SEC with $r_{0}=0.7$ and $M=1$ for $a=\{0.2,0.4,0.6\}$; this panel shows the delimited region of SEC violation.}
    \label{fig:SEC}
\end{figure}

Figure~\ref{fig:SEC} summarizes the behavior of the strong energy condition. For fixed $r_{0}$ (left), increasing the string strength $a$ amplifies the negative layer of $\mathrm{SEC}$ and pushes its zero outward, indicating a thicker interior band where the Tolman mass is negative. For fixed $a$ (right), growing $r_{0}$ lifts $\mathrm{SEC}$ and displaces the zero to larger $r$, shrinking that band and smoothing the transition to the string–cloud exterior where $\mathrm{SEC}\!\to\!0^{+}$. The radial profiles (bottom) make this trend explicit: larger $a$ deepens and slightly widens the dip before the asymptotic return to zero. This structure contrasts with de~Sitter–core NED black holes—where SEC often fails already at the center \cite{Bronnikov2001,Dymnikova2004}—and is consistent with Zaslavskii’s integral argument that a static interior must contain a SEC–violating layer between the regular core and the exterior \cite{Zaslavskii2010}.

\begin{figure}[h!]
    \centering
    \includegraphics[width=0.49\linewidth]{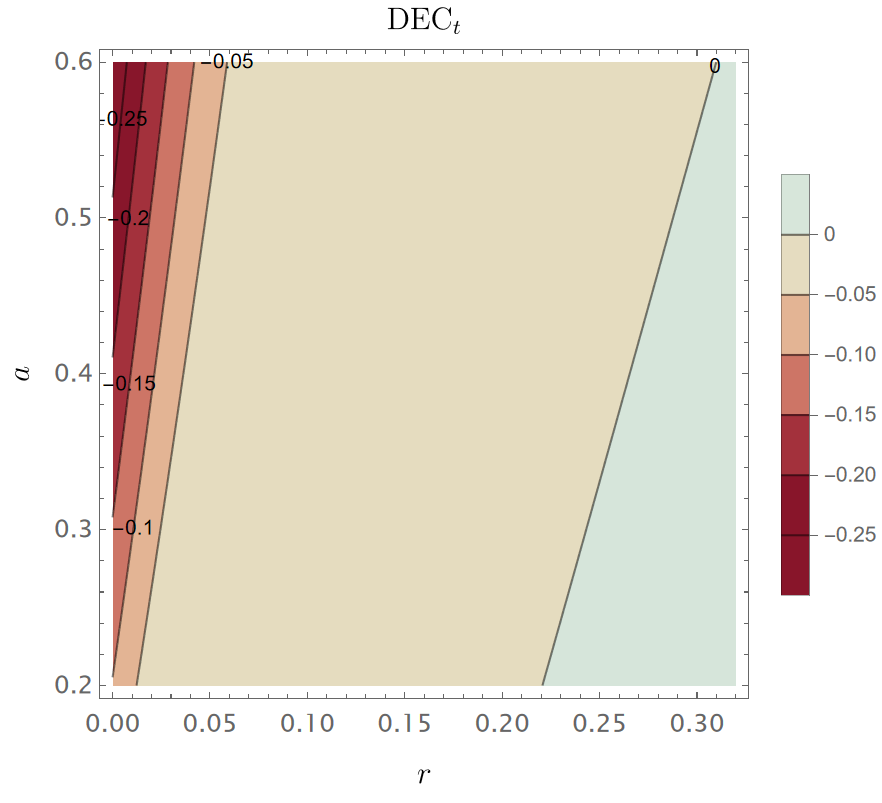}
    \includegraphics[width=0.49\linewidth]{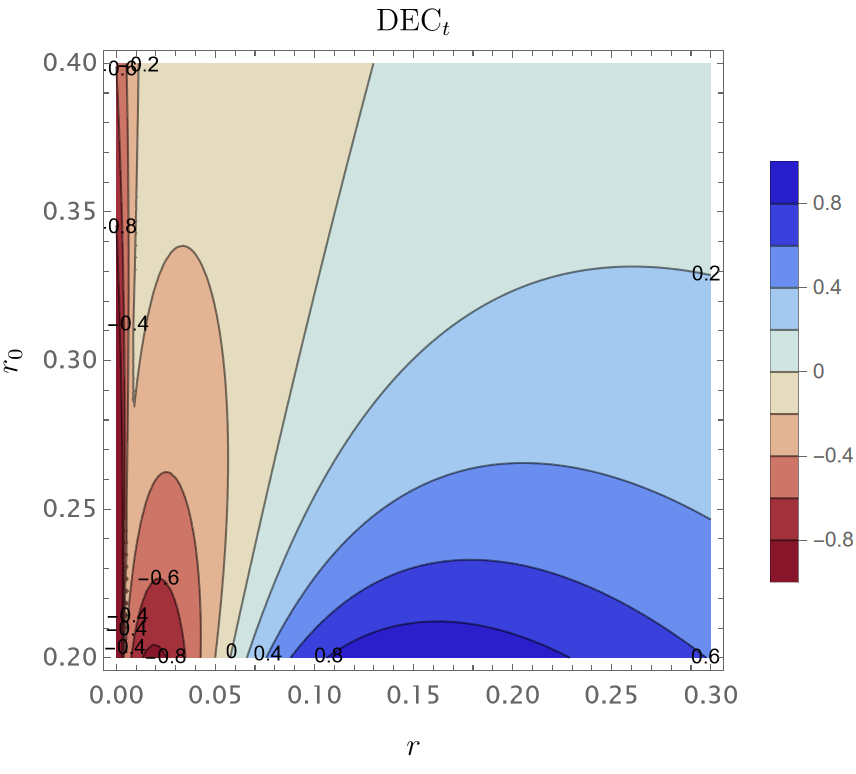}
    \caption{Contours of the tangential DEC ($\text{DEC}_t$). in the {left} panel, we set $r_{0}=0.7$ and $M=1$; larger $a$ shifts the $\mathrm{DEC}_t=0$ line outward and reduces the inner region where $\mathrm{DEC}_t>0$ (DEC satisfied). In the {right} panel, we set $a=0.4$ and $M=1$. The contour $\mathrm{DEC}_t=0$ marks the DEC boundary.}
    \label{fig:DEC}
\end{figure}

Figure~\ref{fig:DEC} displays tangential DEC ($\mathrm{DEC}_t$). For fixed $r_{0}=0.7$ (left), increasing $a$ makes $\mathrm{DEC}_t$ more negative at fixed $r$ and pushes the $\mathrm{DEC}_t=0$ boundary outward, signalling a reduced domain of DEC validity—consistent with the growth of the anisotropic layer seen in $\mathrm{WEC}_t$. For fixed $a=0.4$ (right), enlarging $r_{0}$ barely changes the narrow violation right at $r=0$, but it noticeably \emph{shrinks} the intermediate-$r$ region where $\mathrm{DEC}_t>0$; the $\mathrm{DEC}_t=0$ contour tilts toward smaller $r$, indicating a loss of DEC support as the regularization scale increases. At large $r$ (beyond the plotted window) $\mathrm{DEC}_t\to\rho>0$, in accordance with the string–cloud asymptotics.

\section{Thermodynamics}\label{tmd}

The thermodynamic properties of our regularized black hole can be derived from the metric function introduced in the previous section. Before proceeding, it is necessary to distinguish between the internal and external horizons by examining how the mass depends on the radius of the horizon, as determined from the condition \( f(r_h) = 0 \). Solving for the mass leads to  
\begin{equation}
    M(r_h)=\frac{r_h}{2}+\frac{
4 r_h^3 r_0 + 6 r_h^2 r_0^2 + 4 r_h r_0^3 + r_0^4 +
r_h^2 r_0^2 |a| \, {}_2F_1\!\left(-\tfrac{1}{2}, -\tfrac{1}{4}; \tfrac{3}{4}; -\tfrac{r_h^4}{r_0^4}\right)
}{
2 r_h^3
},\label{M(rh)}
\end{equation}
where the Schwarzschild black hole mass is recovered in the limit \( r_0 \to 0 \). The mass function $M(r_h)$ exhibits a minimum $M_{\rm ext}$ at $r_h=r_h^{\min}$, corresponding to the extremal configuration where the inner and outer horizons coincide ($r_+=r_-=r_h^{\min}$). Thermodynamic quantities (temperature, entropy, heat capacity, free energies) must be referred to the outer horizon $r_+$; therefore, the thermodynamic analysis is meaningful only for parameter ranges yielding a real outer horizon, i.e. $M\ge M_{\rm ext}$ (or equivalently $r_+\ge r_h^{\min}$). Values of $r_h$ below $r_h^{\min}$ correspond either to the inner horizon or to no horizon at all, and therefore do not represent physically admissible event horizons for the equilibrium thermodynamics considered here. The extremal case $r_+=r_h^{\min}$ should be treated separately, since it typically leads to $T\to 0$ and requires an interpretation in terms of remnants. From now on, we will consider $r_+=r_h$, unless otherwise stated.

\begin{figure}[h!]
    \centering 
       \includegraphics[width=0.495\textwidth]{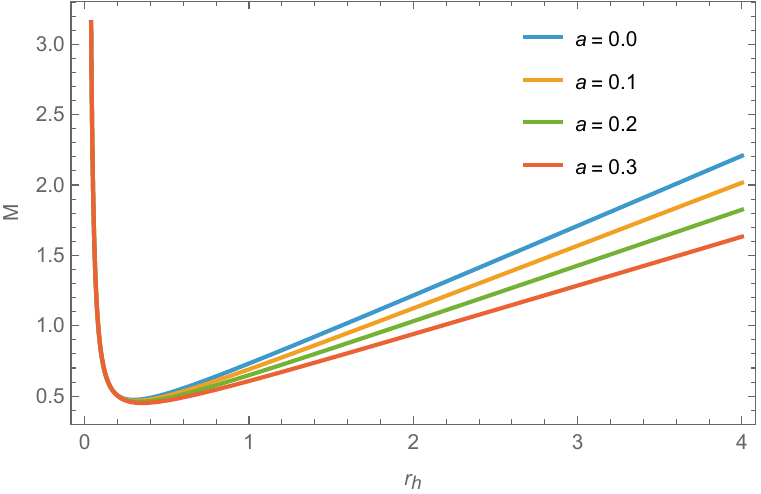}
        \includegraphics[width=0.495\textwidth]{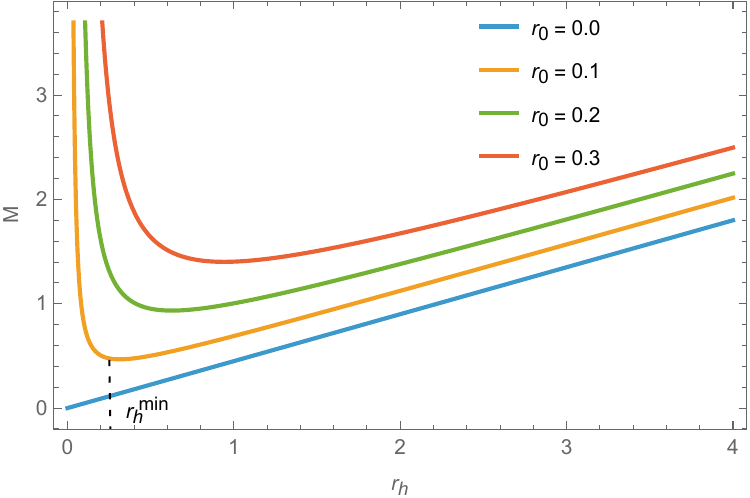}
    \caption{The black hole mass as a function of the event horizon radius $r_h$. 
    In the left panel, the dependence on the string cloud parameter is displayed ($r_0=0.1$), 
    while in the right panel the variation with the regularization parameter is shown ($a=0.1$), with $r_h^{min}\approx 0.3$ for $r_0=0.1$ denoting the lower (upper) bound of the outer (inner) horizon radius. For a fixed event horizon radius, increasing the string cloud parameter decreases the black hole mass, whereas a larger regularization parameter leads to a higher mass.}
    \label{fig:massrh}
\end{figure}

Figure~\ref{fig:massrh} shows the behavior of the black hole mass as a function of \( r_h \). 
As we stated previously, the outer (event) horizon corresponds to the branch with \( r_h > r_h^{\text{min}} \), that is, radii larger than those associated with the minimal mass configuration, which marks the point where the inner and outer horizons merge into a single degenerate horizon. In the left panel, the mass function is shown for different values of the string cloud parameter, while in the right panel it is displayed for a varying regularization parameter, with $r_h^{min}\approx 0.3$ for $r_0=a=0.1$. It is observed that, as the string cloud parameter increases, the black hole mass decreases for a fixed event horizon radius, whereas increasing the regularization parameter leads to a larger black hole mass. When this parameter vanishes, the behavior is that of the Schwarzschild black hole.

\begin{figure}[h!]
    \centering 
       \includegraphics[width=0.495\textwidth]{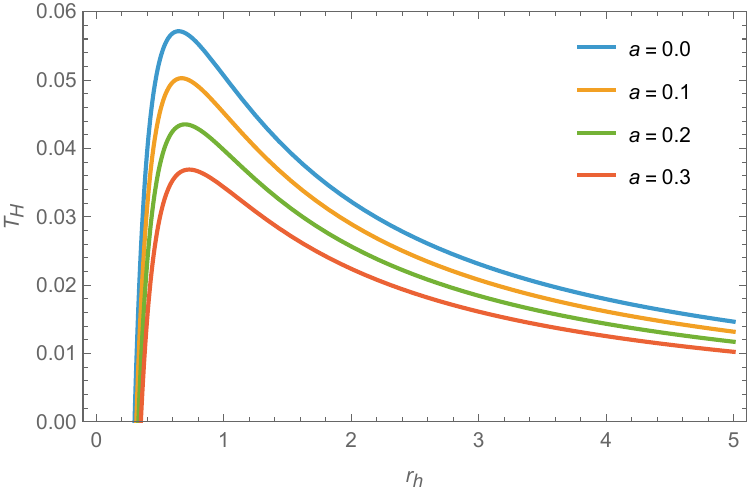}
        \includegraphics[width=0.495\textwidth]{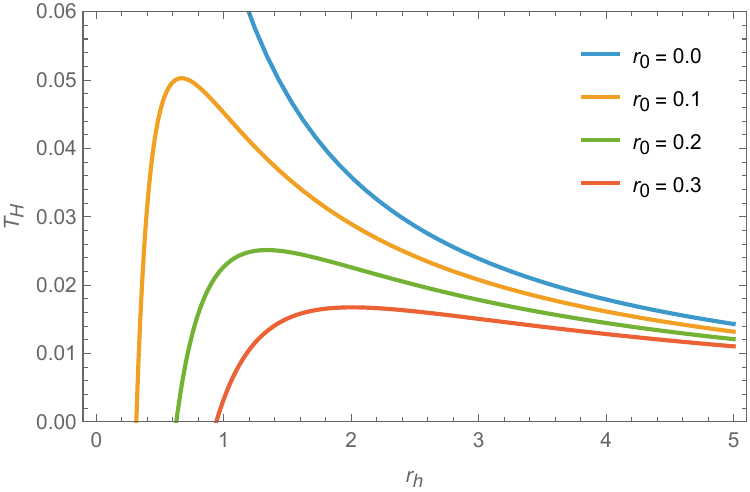}
    \caption{Hawking temperature as a function of the event horizon radius $r_h$. 
    In the left panel, the dependence on the string cloud parameter is displayed ($r_0=0.1$), 
    while in the right panel the variation with the regularization parameter is shown ($a=0.1$). Only positive values of $T_H$ are depicted.}
    \label{fig:temprh}
\end{figure}

Following the standard approach, the Hawking temperature is obtained from the surface gravity at the event horizon according to $T_H = f'(r_h)/(4\pi)$. Applying this definition to the present model yields the simple expression
\begin{equation}
T_H = \frac{1}{4\pi r_h}
\left[
1 - 
\frac{4 r_0 (r_h + r_0)^3 + r_h^2 \sqrt{1 + r_h^4 / r_0^4} \, r_0^2 |a|}
{(r_h + r_0)^4}
\right].
\label{TH}
\end{equation}
This expression explicitly depends on the string cloud parameter \(a\) and the regularization scale \(r_0\). 
The combined effect of these parameters modifies the temperature profile, lowering it relative to that of a Schwarzschild black hole, \(T_{\rm Schw} = (4\pi r_h)^{-1}\), which is recovered in the limit \(r_0 \to 0\). This modification enables the emergence of critical points where the black hole undergoes transitions between distinct thermodynamic phases. Fig.~\ref{fig:temprh} shows the Hawking temperature as a function of the event horizon radius, for different values of the string parameter (left) and the regularization scale (right). In both cases, the temperature increases as \( r_H \) decreases, reaches a maximum, and then drops to zero, indicating a phase transition and the formation of a remnant when $T_H=0$. Furthermore, it is observed that increasing either the string cloud parameter \( a \) or the regularization scale \( r_0 \) shifts the remnant radius to larger values, consequently reducing the temperature peak and facilitating the phase transition process.

Interestingly, although the temperature depends on the string parameter $a$, the entropy does not. In fact, the entropy of the regular black hole, obtained by integrating the first law of thermodynamics $dM = T_H\, dS_{bh}$, reads
\begin{equation}
S_{bh} = \pi
\left[
r_h^2 + 8 r_h r_0 - \frac{8 r_0^3}{r_h} - \frac{r_0^4}{r_h^2}
+ 6 r_0^2 \ln\!\left(\frac{r_h^2}{r_0^2}\right)
\right].
\label{Sbh}
\end{equation}
This expression differs from the standard Bekenstein-Hawking area law due to the logarithmic and inverse-radius corrections introduced by the regularization and string cloud extension scale $r_0$, yet it is remarkable that it remains completely independent of the string cloud parameter $a$. This suggests that the microscopic degrees of freedom responsible for the entropy are primarily associated with the regularized geometry rather than with the external string field. In other words, while the string cloud affects global thermodynamic quantities through the temperature and mass distribution, it does not directly contribute to the counting of horizon microstates, whose structure is governed by the core regularization parameter $r_0$.

In regard to the heat capacity of the black hole, which can be computed via \(
   \displaystyle  C=\frac{dM}{dT_H}=\frac{dM/dr_h}{dT_H/dr_h},
\)
it follows from (\ref{M(rh)}) and (\ref{TH}) that
\begin{figure}[h!]
    \centering 
       \includegraphics[width=0.496\textwidth]{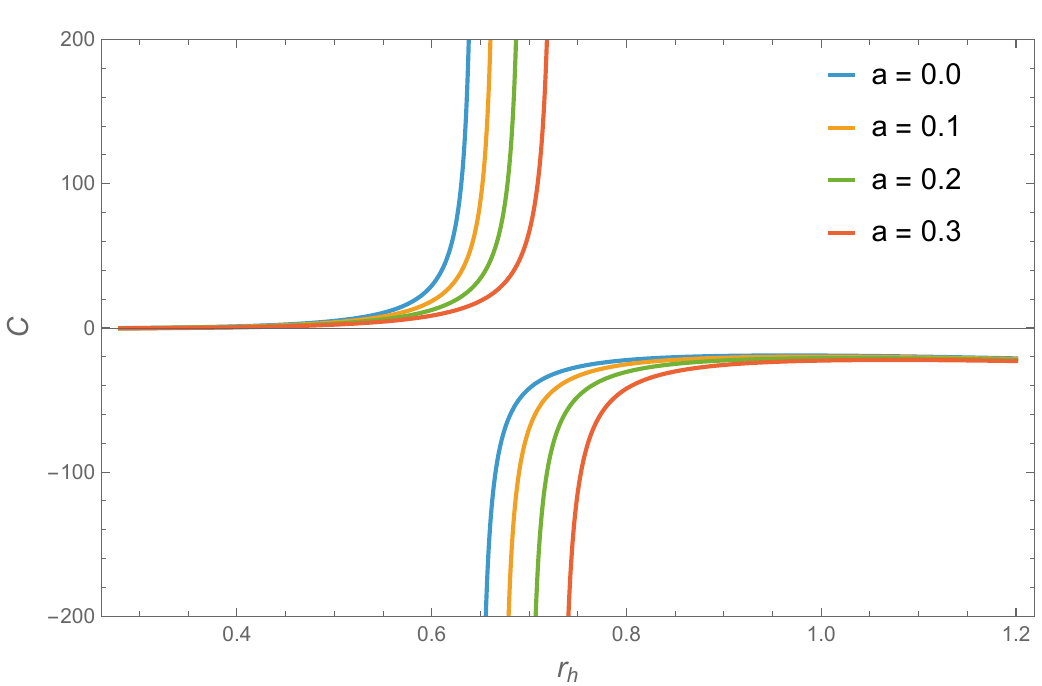}
        \includegraphics[width=0.496\textwidth]{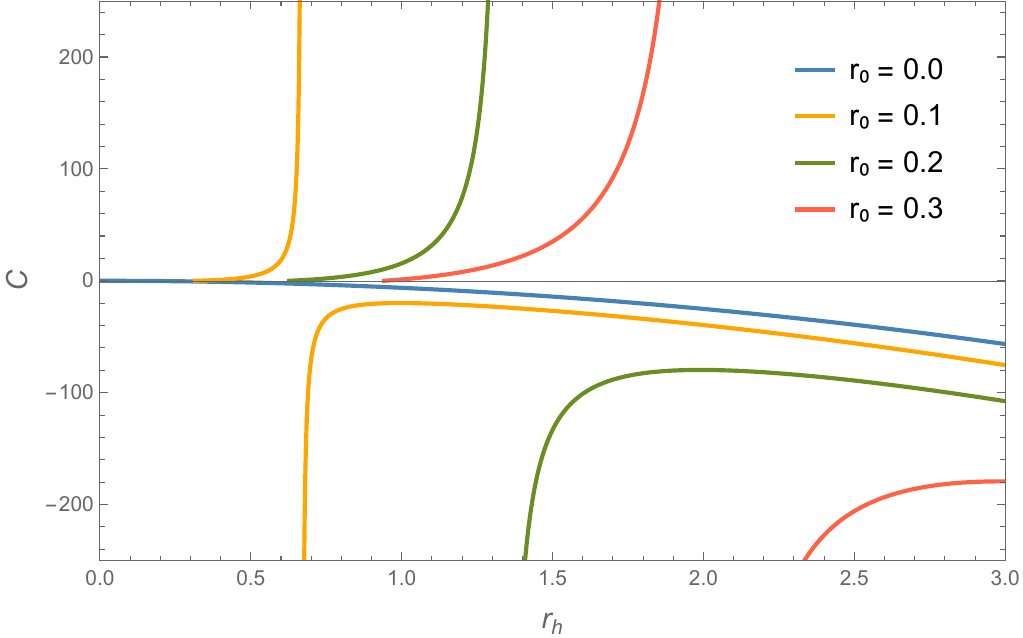}
    \caption{Heat capacity as a function of the event horizon radius $r_h$. In the left panel, the dependence on the string cloud parameter $(r_0 = 0,1)$ is shown, while in the right panel the variation with the regularization parameter $(a = 0,1)$ is displayed, with the blue curve indicating the Schwarzschild black hole.}
    \label{fig:Caprh}
\end{figure}
\begin{equation}
\displaystyle C(r_h) = -{2 \pi (r_0 + r_h)^2}\left[\frac{(3r_0 - r_h)(r_0 + r_h)^3 \sqrt {r_0^4+{r_h^4 }} + r_h^2(r_0^4 + r_h^4)\, |a|
}{r_h^2 (3 r_0^2 + 6 r_0 r_h - r_h^2) \sqrt{r_0^4+r_h^4} + r_h^4 \displaystyle \frac{r_h^5- 3 r_0 r_h^4+3 r_0^4 r_h-r_0^5}{(r_0+r_h)^3}\, |a|
}\right],\label{C(rh)}
\end{equation}
where the heat capacity of the Schwarzschild black hole, \(C_{\rm Schw} = -2\pi r_h^2\), is recovered in the limit \(r_0 \to 0\). 
The behavior of the heat capacity as a function of the event horizon radius, for different values of the string parameter (left) and the regularization scale (right), is shown in Fig.~\ref{fig:Caprh}. In both cases, variations in the string parameter \(a\) and in the regularization scale \(r_0\) shift the location of the phase transition, thereby altering the boundaries between local, thermodynamically unstable and stable regimes, manifested by a discontinuous transition from negative to positive values, respectively, as the radius of the event horizon decreases. This feature indicates therefore a second-order phase transition, which will be corroborated by the topological thermodynamic analysis discussed in the following section.

To further investigate possible non-extensive effects in black hole thermodynamics, we employ the Rényi entropy formalism. This entropy, which extends the non-extensive Tsallis entropy, introduces a deformation parameter $\lambda$ that quantifies deviations from additivity and is defined as
\begin{equation}
S_{R} = \frac{1}{\lambda}
\log{\left(1 + \lambda S_{bh}\right)},
\label{SR}
\end{equation}
where the standard black hole entropy $S_{bh}$, given by Eq. (\ref{Sbh}), is recovered in the limit $\lambda\to 0$. This entropy provides a consistent extension of the Boltzmann–Gibbs formalism for systems exhibiting correlations, memory effects, or long-range interactions that violate extensivity \cite{Tsallis:1987eu,Renyi:1959pbs}. In gravitational contexts, the Rényi framework allows one to treat black holes as non-extensive thermodynamic systems while preserving key properties such as concavity and stability \cite{Czinner:2015eyk,Czinner:2017tjq}. It has been successfully applied to model quantum and statistical corrections to the horizon entropy, offering an effective description of deviations from classical thermodynamic behavior in gravitational systems \cite{Biro:2013cra}. 

Moreover, the Rényi entropy framework is also suitable for systems with a finite or small number of degrees of freedom, where standard thermodynamic extensivity breaks down. In such cases, statistical fluctuations and correlations among microstates become non-negligible, and the Rényi formalism provides an effective way to incorporate these non-extensive effects while maintaining a well-defined temperature and equilibrium condition \cite{Jizba:2002um}.

The Rényi temperature of the black hole can be calculated via
\begin{equation}
\begin{split}
T_R
&= \frac{dM}{dS_R}
= T_H(1+\lambda S_{bh})  \\
&= 
\left[\frac{
\left(r_h - 3 r_0\right)(r_h + r_0)^{3}
-
r_h^{2}\sqrt{1 + \frac{r_h^{4}}{r_0^{4}}}\, r_0^{2}\, |a|
}{4\pi r_h (r_h + r_0)^{4}}\right]
\left[
1
+ \frac{\pi \lambda \left(r_h^{4} + 8 r_h^{3} r_0 - 8r_h r_0^{3} - r_0^{4}\right)}{r_h^{2}}
+ 6\pi r_0^{2}\lambda \ln\!\left(\frac{r_h^2}{r_0^2}\right)
\right],
\end{split}
\end{equation}
where we can note that it is always greater than the corresponding Hawking temperature, provided $\lambda>0$.

\begin{figure}[h!]
    \centering 
       \includegraphics[width=0.496\textwidth]{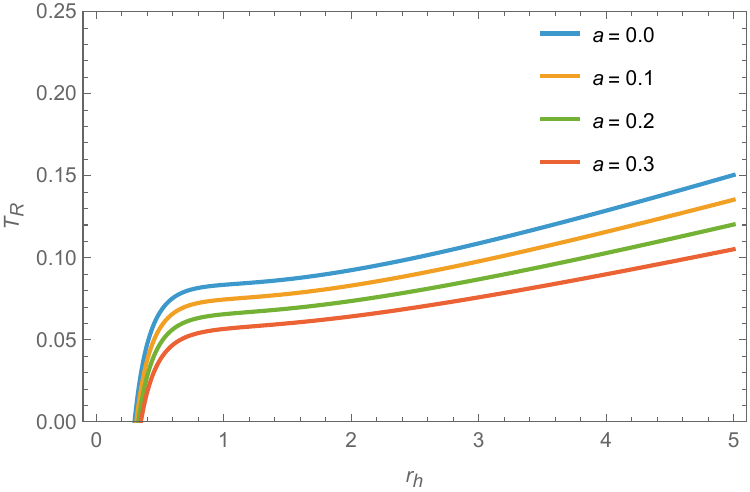}
        \includegraphics[width=0.496\textwidth]{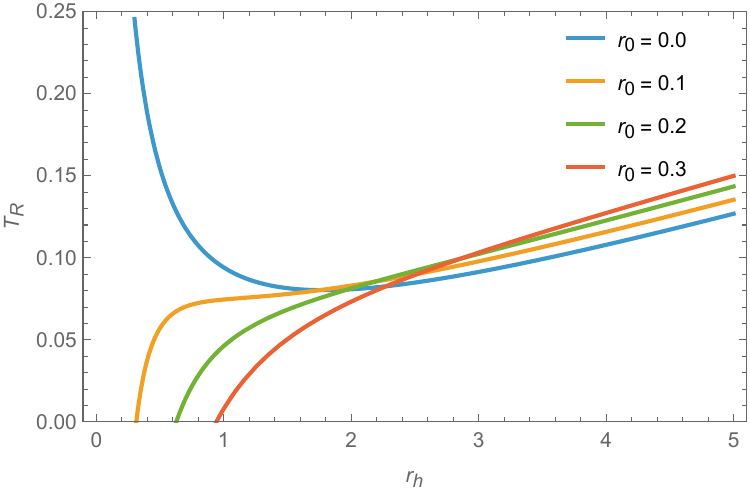}
    \caption{Rényi temperature, for $\lambda=0.1$, as a function of the event horizon radius $r_h$. In the left panel, the dependence on the string cloud parameter $(r_0 = 0,1)$ is shown, while in the right panel the variation with the regularization parameter $(a = 0,1)$ is displayed, with the blue curve indicating the Schwarzschild black hole.}
    \label{renyitemp}
\end{figure}

Figure~\ref{renyitemp} shows the Rényi temperature as a function of the event horizon radius. For the regular black holes supported by the string cloud, the curves exhibit no local stationary points, indicating that they do not undergo any phase transition, contrary to what happens in the standard Gibbs--Boltzmann thermodynamics. In contrast, the Schwarzschild case (\(r_0 \to 0.0\)) does display a local minimum, as illustrated in the right panel of the figure.

Moreover, the behavior of the Rényi temperature confirms the existence of remnants, except for the Schwarzschild black hole.  

\begin{figure}[h!]
    \centering 
       \includegraphics[width=0.493\textwidth]{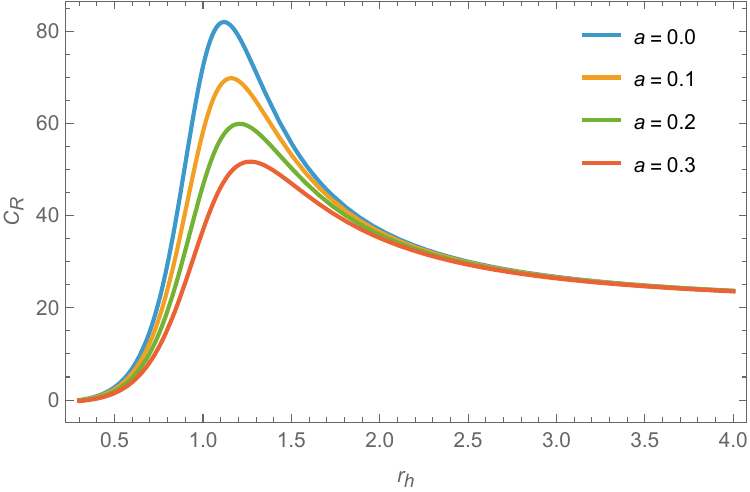}
        \includegraphics[width=0.5\textwidth]{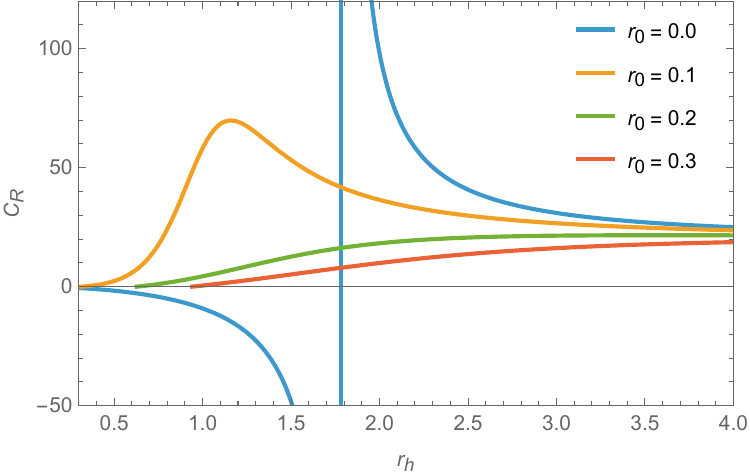}
    \caption{Rényi heat capacity, for $\lambda=0.1$, as a function of the event horizon radius $r_h$. In the left panel, the dependence on the string cloud parameter $(r_0 = 0,1)$ is shown, while in the right panel the variation with the regularization parameter $(a = 0,1)$ is displayed, with the blue curve indicating the Schwarzschild black hole.}
    \label{renyiheat}
\end{figure}

The absence of phase transitions in the regular black holes under consideration and their local thermodynamic stability is corroborated through the behavior of the Rényi heat capacity, defined as $C_{R} = dM/dT_{R}$. This is precisely what Fig.~\ref{renyiheat} illustrates: while the regular black hole supported by the string cloud exhibits no signatures of thermodynamic instability, the Schwarzschild black hole again displays a characteristic phase-transition point. All these results will be further confirmed in the next section through the analysis of topological thermodynamics.

\section{Black holes as topological defects}\label{bhtd}

Recently, Wei \emph{et al.}~\cite{wei2022black} showed that black holes can be regarded as topological defects. 
They introduced a off-shell generalized Gibbs free energy
\begin{equation}\label{GFE}
    \mathcal{F} = M - \frac{S}{\tau},
\end{equation}
with $\tau > 0$, and defined a two-dimensional vector field
\begin{equation}\label{phi}
    \vec{\phi} = \left( \frac{\partial \mathcal{F}}{\partial r_h},\,-\cot{\Theta}\csc{\Theta} \right).
\end{equation}
The zeros of $\vec{\phi}$ occur at $\Theta = \pi/2$ and $\tau = 1/T$.  
Associated with $\vec{\phi}$, they introduced a topological current
\begin{equation}\label{j}
    j^\mu = \frac{1}{2\pi}\,\epsilon^{\mu\nu\rho}\epsilon_{ab}\,\partial_\nu n^a\,\partial_\rho n^b,
\end{equation}
where $\mu,\nu,\rho = 0,1,2$, and $a,b = 1,2$. The unit vector $\vec{n}$ is defined as
\begin{equation}\label{n}
    n^1 = \frac{\phi^1}{\phi}, \quad n^2 = \frac{\phi^2}{\phi}, \quad \phi = |\vec{\phi}|.
\end{equation}
The current (\ref{j}) is algebraically conserved and vanishes everywhere except at the zeros of $\vec{\phi}$~\cite{wei2022black}.  
From Eq.~(\ref{j}), one obtains
\begin{equation}
    j^0 = \frac{1}{\pi}\left( \partial_1 n^1 \partial_2 n^2 - \partial_2 n^1 \partial_1 n^2 \right).
\end{equation}

\begin{figure}[h!]
    \centering 
    \includegraphics[width=0.48\textwidth]{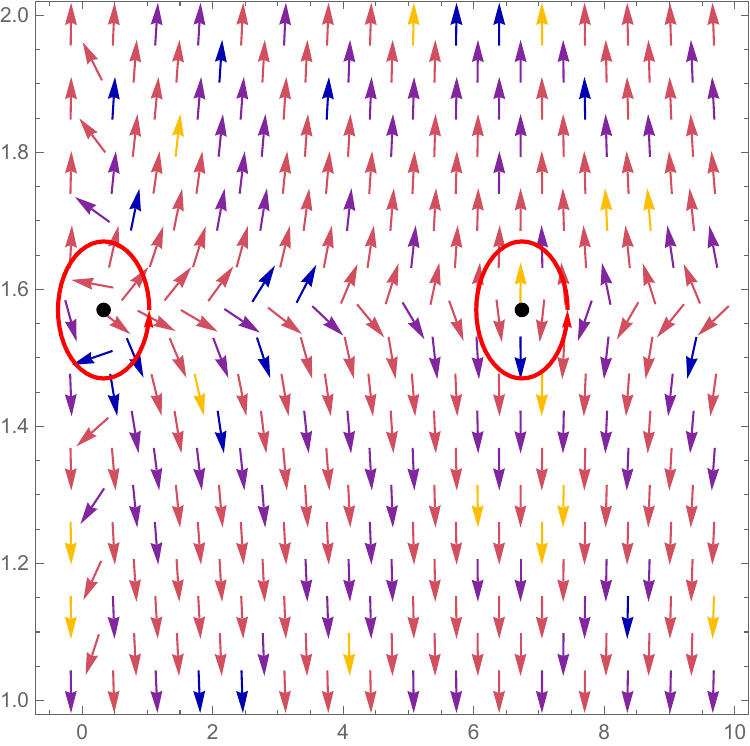}
    \includegraphics[width=0.48\textwidth]{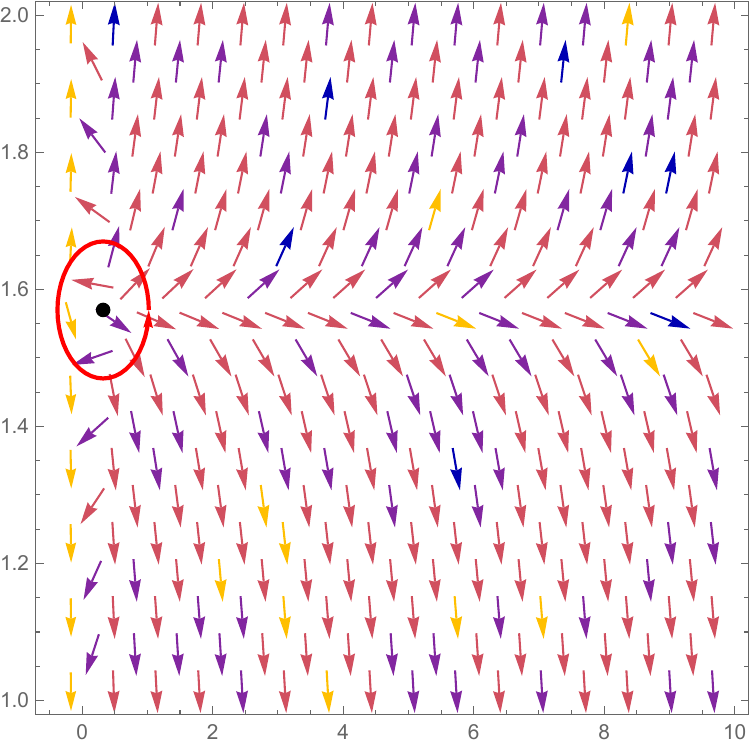}
    \includegraphics[width=0.49\textwidth]{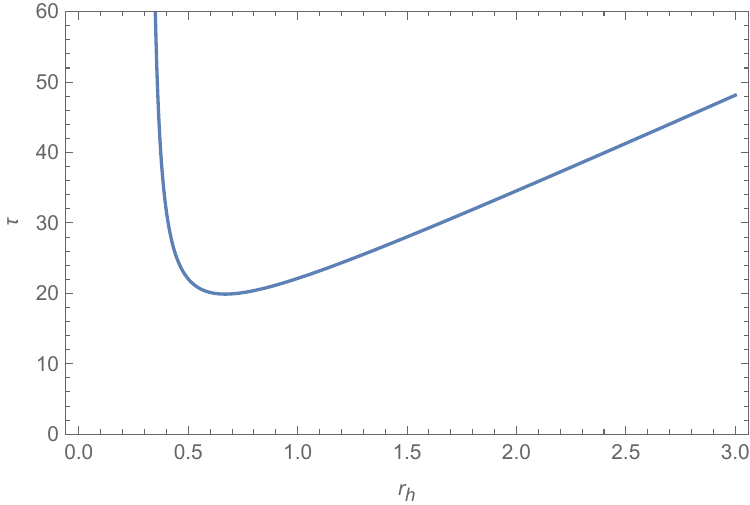}
    \includegraphics[width=0.49\textwidth]{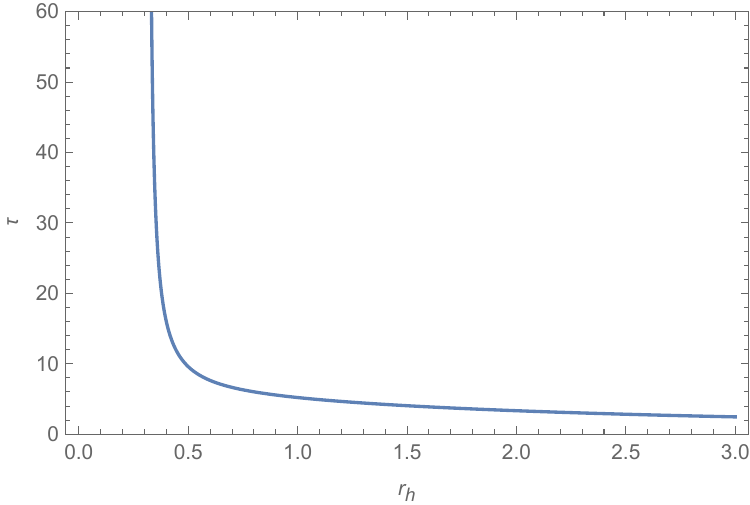}
    \caption{Top panels: Phase-space vector field \((r_h, \Theta)\) for \(\tau = 100\). Left: With standard black hole entropy (\(\lambda = 0\)), showing two states -- stable (\(W=+1\)) and unstable (\(W=-1\)) -- separated by a phase transition. Right: With Rényi entropy (\(\lambda = 0.5\)), exhibiting a single stable microscopic state (\(W=+1\)). Bottom panels: Respective \(\tau(r_h)\) parameter at the zero points. Left: With a local minimum at the transition point. Right: Without stationary points, confirming the absence of phase transition. Set parameters: \(r_0 = a = 0.1\).}
    \label{fig:param_space2}
\end{figure}

For regions where $\phi \neq 0$ and $\phi^a$ are smooth functions, this can be rewritten as
\begin{equation}
    j^0 = \frac{1}{\pi}\left( \partial_1 Q - \partial_2 P \right), 
    \quad Q = n^1 \partial_2 n^2, \quad P = n^1 \partial_1 n^2.
\end{equation}
\begin{figure}[h!]
    \centering 
      \includegraphics[width=0.7\linewidth]{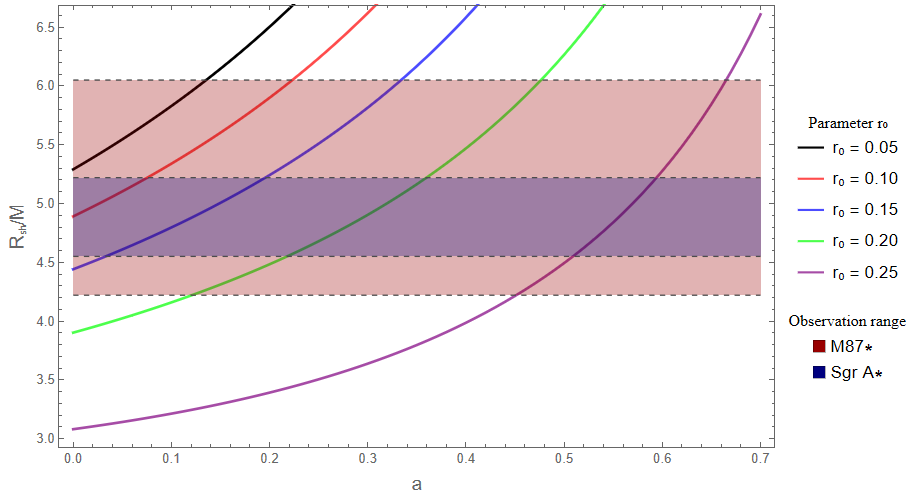}
    \includegraphics[width=0.6\textwidth]{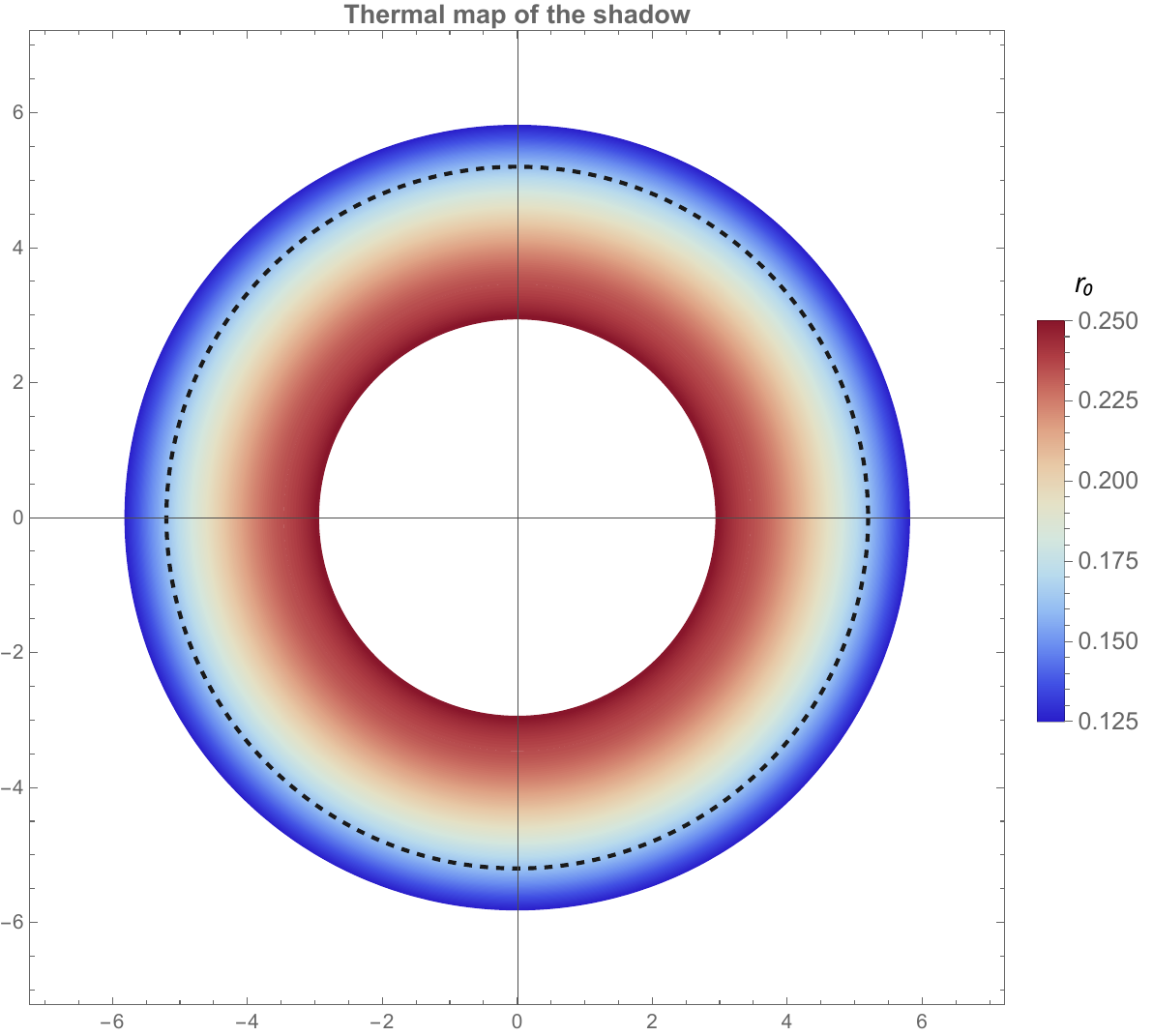}
    \caption{Top panel: Shadow radius $R_{\mathrm{sh}}/M$ as a function of the cloud parameter $a$ for different values of the regularization parameter $r_0$. The colored curves represent the theoretical predictions from the photon sphere condition, while the shaded regions correspond to the observational ranges of Sgr~A* (purple band, $4.55 \le R_{\mathrm{sh}}/M \le 5.22$) and M87* (purple plus pink bands, $4.22 \le R_{\mathrm{sh}}/M \le 6.05$), as reported by the EHT Collaboration. Bottom panel: Polar thermal map of the shadow radius, showing the continuous variation with $r_0$ ($0.125 \le r_0 \le 0.250$), for $a=0.3$; the dashed circle marks the Schwarzschild shadow limit.}
    \label{fig:shadowplots}
\end{figure}
Hence, for a domain $D$ where $\phi \neq 0$, the Green’s theorem yields
\begin{align*}
  0 &= \int_D j^0 \, d^2x 
     = \frac{1}{\pi} \int_D (\partial_1 Q - \partial_2 P) \, d^2x 
     = \frac{1}{\pi} \oint_\Sigma (P \, dx^1 + Q \, dx^2)
     = \frac{1}{\pi} \oint_\Sigma n^1 \, dn^2.
\end{align*}
For a closed contour $C$ enclosing all zeros of $\vec{\phi}$, the corresponding topological number is defined as
\begin{equation}\label{W}
   W = \frac{1}{\pi} \oint_C n^1 \, dn^2 
     = \frac{1}{\pi} \sum_{i=1}^{N} \oint_{c_i} n^1 \, dn^2,
\end{equation}
where each $c_i$ is an arbitrary small contour surrounding a zero of $\vec{\phi}$.

In Fig.~\ref{fig:param_space2}, the phase-space dynamics of the black hole are illustrated for both the standard and Rényi entropies. For the standard case (\(\lambda = 0\)), shown in the left panels, two distinct black hole states appear in the phase space: a stable configuration (\(W = +1\)) and an unstable one (\(W = -1\)), which are separated by a phase transition. This behavior is reflected in the profile of \(\tau(r_h)\), where a local minimum marks the transition point between the two states. In contrast, for the Rényi entropy (\(\lambda = 0.5\)), displayed in the right panels, the system exhibits only one stable configuration (\(W = +1\)), and the corresponding function \(\tau(r_h)\) becomes monotonic, without stationary points. This monotonic behavior confirms the absence of a phase transition, indicating that the non-extensive parameter \(\lambda\) smooths out the thermodynamic landscape and enhances the stability of the black hole configuration. Such a feature is consistent with the intrinsic non-additive character of the Rényi formalism, which effectively restricts the number of accessible microstates and favors a microscopic regime with fewer active degrees of freedom.

\section{Shadow analysis}\label{sa}

While shadow analyses are often more relevant for rotating geometries, we focus here on the static configuration in order to isolate and systematically evaluate the impact of the string cloud parameters on the shadow morphology. This simplified setup represents a crucial first step toward confronting the present regular black hole model with observational data, which have in turn motivated a number of recent investigations \cite{Liu2019pov,Kimet2021,CIMDIKER2021100900,Okyay_2022,PANTIG2023169197,LIU2024139052,Jiang_2024,vgn2025}. In this section, we estimate the black hole shadow radius and establish preliminary bounds on the free parameters of our model. For computational tractability and to isolate the fundamental geometric effects, we neglect any contribution from accretion flows or plasma and assume a distant observer located in the asymptotic region.

For a static, spherically symmetric spacetime, the boundary of the shadow as seen by a distant observer is determined by the critical photon sphere. The radius of the photon sphere \( r_{\mathrm{ph}} \) is defined by the existence of unstable circular null orbits. A quite useful approach to characterizing these orbits is through the function \cite{Perlick:2021aok}
\begin{equation}  \label{gamma}
    \gamma(r) = \sqrt{-\frac{g^{tt}(r)}{g^{\phi\phi}(r)}},
\end{equation}
which corresponds to the impact parameter \( b = L/E \) of a photon orbiting at radius \( r \). The critical radius \( r_{\mathrm{ph}} \) is then found by solving the extremization condition
\begin{equation}  \label{r_ph}
    \frac{d\gamma^2(r)}{dr} = 0 \quad \text{at} \quad r = r_{\mathrm{ph}}.
\end{equation}
The corresponding critical impact parameter is \( b_{\mathrm{ph}} = \gamma(r_{\mathrm{ph}}) \), which separates photons that fall into the black hole from those that escape to infinity. For an observer at a finite distance \( R_o \), the observed shadow radius \( R_{\mathrm{sh}} \) is related to the critical impact parameter by \cite{Perlick:2021aok}
\begin{equation}  \label{Rsh_finite}
    R_{\mathrm{sh}} = b_{\mathrm{ph}} \sqrt{ - g_{tt}(R_o) }.
\end{equation}
This expression incorporates the gravitational redshift effect. In the asymptotic limit \( R_o \to \infty \), the shadow radius reduces to \(R_{\mathrm{sh}} = b_{\mathrm{ph}}(1-|a|)^{1/2}\).

Figure~\ref{fig:shadowplots} summarizes the results of our phenomenological analysis. The top panel shows the dependence of \(R_{\mathrm{sh}}/M\) on the cloud parameter \(a\) for several values of the regularization scale \(r_0\), highlighting the regions of consistency with the observed shadow sizes of Sgr~A* and M87*. The bottom panel presents a thermal map of the shadow morphology, illustrating the continuous variation of \(R_{\mathrm{sh}}\) with \(r_0\) and its deviation from the Schwarzschild shadow limit (dashed circle). Increasing \(r_0\) leads to a smaller shadow radius, as the regular core softens the central curvature and pushes the photon sphere inward. The classical Schwarzschild shadow \(R_{\mathrm{sh}} = 3\sqrt{3}\,M\) is represented by the dashed circle.

\section{Conclusions}\label{conclusion}

In this work, we constructed a new family of regular black hole solutions supported by the Letelier–Alencar string cloud and regularized it through a rational Dagum-type distribution, a regulator that has also been used in the literature to generate regular geometries sourced by nonlinear electrodynamics. The employed regularization successfully smooths the matter profile and ensures the finiteness of curvature invariants, overcoming the limitations of the standard exponential regularization adopted in previous extensions of the Letelier model. The resulting geometry interpolates between a string–cloud–dominated regime at large distances and an anti--de Sitter vacuum core at the origin, characterized by an effective curvature radius determined by the string parameter $a$ and the regularization scale $r_{0}$. In addition, we have shown that our regularized string–cloud black hole and the Frolov regular black hole solution exhibit the same near-origin behavior up to $\mathcal{O}(r^{4})$, as demonstrated by the matching leading terms in the expansions of the metric function and curvature invariants. Although this local agreement confirms that both geometries share the same regular core structure -- AdS-type -- it does not imply that their physical sources are equivalent.

We analyzed in detail the matter source associated with our solution, showing from the time component of the field equations that the resulting energy density profile captures how the interplay between the string cloud strength and the regularization scale governs the transition from the AdS-like inner region to the asymptotically flat exterior dominated by the string cloud. By examining the remaining components of Einstein’s equations, we evaluated the full set of energy conditions and identified the regions where the null, weak, dominant, and strong conditions are satisfied or violated, in accordance with the expected behavior of regular black holes possessing a smooth central core.

A complete thermodynamic analysis was performed. We derived the mass, Hawking temperature, entropy, and heat capacity of the solution, demonstrating in particular that while both $a$ and $r_{0}$ influence the temperature and mass distribution, the black hole entropy, where the Hawking-Bekenstein area law is modified by logarithmic and inverse-radius corrections introduced by the regularization scale $r_0$, depends solely on this scale. This indicates that the microscopic degrees of freedom associated with the entropy are governed by the regular core rather than by the external string field. We also studied the non-extensive thermodynamics of the system through the R\'enyi entropy formalism, showing that the deformation parameter $\lambda$ smooths out the phase structure and eliminates the standard phase transition, yielding a single thermodynamically stable branch. These results were corroborated by the topological thermodynamics approach, which identifies black hole states as topological defects and confirms the absence of phase transitions in the non-extensive regime.

Finally, we examined the observational phenomenology of the solution by computing the shadow radius and establishing constraints on the parameters $a$ and $r_{0}$ from the Event Horizon Telescope measurements of Sgr~A* and M87*. The presence of the regular core reduces the shadow radius with respect to the Schwarzschild case, in agreement with trends observed in other regular black hole models. For realistic parameter ranges, the predicted shadow sizes remain compatible with current observational bounds.

In summary, the regularized string cloud black hole presented here constitutes a consistent and physically motivated extension of recent developments in regular geometries, combining mathematical regularity, well-defined thermodynamic behavior, and phenomenological viability. Future investigations may explore broader theoretical settings, possible generalizations of the matter sector as well as refinements and extensions of the phenomenological analysis, offering additional insight into the role of regular cores and extended matter fields in gravitational physics.

\section*{Acknowledgments}
CRM would like to thank Conselho Nacional de Desenvolvimento Cient\'{i}fico e Tecnol\'ogico (CNPq) for the partial financial support, through grant 301122/2025-3. FBL is funded by Fundação Cearense de Apoio ao Desenvolvimento Científico e Tecnológico (FUNCAP) and by  Conselho Nacional de Desenvolvimento Científico e Tecnológico (CNPq), grant number 305947/2024-9. 

\bibliographystyle{apsrev4-1}
\bibliography{ref.bib}

\end{document}